\documentclass[twocolumn]{aastex63}
\usepackage{graphicx}	% Including figure files
\usepackage{amsmath}	% Advanced maths commands
\usepackage{amssymb}	% Extra maths symbols
\usepackage{epstopdf}
\usepackage{multirow}
\received{2024 Oct. 23}
\shorttitle{{\it IXPE} Mrk 421 Campaign}
\shortauthors{Maksym et al.}
\graphicspath{{./}{figures/}}
\defcitealias{whittle04}{WW04}
\defcitealias{whittle05}{W05}
\defcitealias{fischer13}{F13}
\defcitealias{fischer11}{F11}
\defcitealias{revalski21}{R21}

\begin{document}

\title{A Two-Week {\it IXPE} Monitoring Campaign on Mrk 421}

\newcommand\T{\rule{0pt}{2.6ex}}       % Top strut
\newcommand\B{\rule[-1.2ex]{0pt}{0pt}} % Bottom strut

\correspondingauthor{W. P. Maksym}
\email{walter.p.maksym@nasa.gov}

\author[0000-0002-2203-7889]{W. Peter Maksym}
\affiliation{NASA Marshall Space Flight Center, Huntsville, AL 35812, USA}

\author[0000-0001-9200-4006]{Ioannis Liodakis}
\affiliation{NASA Marshall Space Flight Center, Huntsville, AL 35812, USA}
\affiliation{Institute of Astrophysics, Foundation for Research and Technology-Hellas, GR-70013 Heraklion, Greece}

\author[0000-0001-7163-7015]{M. Lynne Saade}
\affiliation{Science \& Technology Institute, Universities Space Research Association, 320 Sparkman Drive, Huntsville, AL 35805, USA}
\affiliation{NASA Marshall Space Flight Center, Huntsville, AL 35812, USA}

\author[0000-0001-5717-3736]{Dawoon E. Kim}
\affiliation{INAF Istituto di Astrofisica e Planetologia Spaziali, Via del Fosso del Cavaliere 100, 00133 Roma, Italy}
\affiliation{Dipartimento di Fisica, Universit\'{a} degli Studi di Roma “La Sapienza”, Piazzale Aldo Moro 5, 00185 Roma, Italy}
\affiliation{Dipartimento di Fisica, Universit\'{a} degli Studi di Roma ``Tor Vergata", Via della Ricerca Scientifica 1, 00133 Roma, Italy}
%email: dawoon.kim@inaf.it	  

\author[0000-0001-9815-9092]{Riccardo Middei}
\affiliation{INAF Istituto di Astrofisica e Planetologia Spaziali, Via del Fosso del Cavaliere 100, 00133 Roma, Italy}
%email: riccardo.middei@ssdc.asi.it	  

\author[0000-0002-5614-5028]{Laura Di Gesu}
\affiliation{ASI - Agenzia Spaziale Italiana, Via del Politecnico snc, 00133 Roma, Italy
}
%email: laura.digesu@est.asi.it

\author[0000-0001-6314-9177]{Sebastian Kiehlmann}
\affiliation{Institute of Astrophysics, Foundation for Research and Technology-Hellas, GR-70013 Heraklion, Greece}
\affiliation{Department of Physics, University of Crete, GR-70013 Heraklion, Greece}
%skiehl@physics.uoc.gr

\author{Gabriele Matzeu}
\affiliation{G. A. Matzeu - Quasar Science Resources SL for ESA, European Space Astronomy Centre (ESAC), Science Operations Department, 28692, Villanueva de la Ca\~{n}ada, Madrid, Spain}
%email:gabriele.matzeu@ext.esa.int}

\author[0000-0002-3777-6182]{Iv\'{a}n Agudo}
\affiliation{Instituto de Astrof\'{i}sica de Andaluc\'{i}a, IAA-CSIC, Glorieta de la Astronom\'{i}a s/n, 18008 Granada, Spain}
%email: iagudo@iaa.es

\author[0000-0001-7396-3332]{Alan P. Marscher}
\affiliation{Institute for Astrophysical Research, Boston University, 725 Commonwealth Avenue, Boston, MA 02215, USA}
%email: marscher@bu.edu

% ...

%% deprecated in this version as it is no longer necessary. AASTeX 



%%%%%%%%%%%%%%%%%
\begin{abstract}
X-ray polarization is a unique new probe of the particle  acceleration in astrophysical jets made possible through the Imaging X-ray Polarimetry Explorer. Here we report on the first dense X-ray polarization monitoring campaign on the blazar Mrk~421. Our observations were accompanied by an even denser radio and optical polarization campaign. We find significant short-timescale variability in both X-ray polarization degree and angle, including a $\sim90^\circ$ angle rotation about the jet axis. We attribute this to random variations of the magnetic field, consistent with the presence of turbulence but also unlikely to be explained by turbulence alone. At the same time, the degree of lower-energy polarization is significantly lower and shows no more than mild variability. Our campaign provides further evidence for a scenario in which energy-stratified shock-acceleration of relativistic electrons, combined with a turbulent magnetic field, is responsible for optical to X-ray synchrotron emission in blazar jets.
\end{abstract}
%%%%%%%%%%%%%%%%%



%%%%%%%%%%%%%%%%%%%%%%%%%%%%%%%%%%%%%%%%%%%%%%%%%%%%%%%%%%%%%%%%%%%%%%%%

\section{Introduction} \label{sec:intro}

Blazars are powerful active galactic nuclei (AGN) with relativistic jets emitting across the electromagnetic (and possibly high-energy particle) spectrum  \citep[e.g.,][]{Blandford2019,Hovatta2019}. Their extreme brightness and variability are attributed to relativistic effects dominating their multi-wavelength emission due to orientation of the jets toward the line of sight \citep[e.g.,][]{Liodakis2018}.  The broad spectral energy distribution (SED) of high-synchrotron-peaked (HSP, $\nu_{syn}>10^{15}$ Hz) blazars is characterized by a synchrotron component extending from radio to X-rays, plus a high-energy component from X-rays to TeV $\gamma$-rays of currently debated origins. 

Since its launch in December 2021, the Imaging X-ray Polarimetry Explorer \cite[IXPE,][]{Weisskopf2022_ixpe_technical} has been shaping our view of high-energy processes in the Universe, including jet-disk geometry as well as particle acceleration and emission in astrophysical jets from accreting black-hole systems \cite[e.g.,][]{Krawczynski2022,Ehlert2022,Marshall2024}. The first {\it IXPE} observation of the blazar Mrk 501 measured a factor of $\sim2$ higher X-ray polarization than at optical or radio wavelengths, with the electric-vector position angles (EVPAs) at all sampled wavelengths roughly aligned with the jet axis on the sky \citep{Liodakis2022nature}, as measured with Very Long Baseline Array (VLBA) imaging \cite[e.g.,][]{Weaver2022}. Synchrotron radiation with such an EVPA arises in a plasma with a mean magnetic field direction aligned perpendicular to the jet axis, as expected for a shock-compressed or helical (or toroidal) field. The observed degree of polarization was a factor $\gtrsim7$ lower than the maximum value of 70-75\% and time-variable, both of which are characteristics of a turbulent magnetic field \citep{Marscher2014}. The presence of a shock can explain how the radiating electrons are accelerated. Radiative energy losses as the electrons propagate away from the shock front lead to energy stratification that causes the emission region to be confined closer to the shock front at higher photon energies \citep{Marscher1985,Marscher2014,Angelakis2016,Tavecchio2018}. The emission regions at different frequencies are therefore, at most, only partially co-spatial, with the larger, lower-frequency regions containing a greater number of cross-polarized turbulent cells, which causes their polarization to be lower. Subsequent {\it IXPE} observations of blazars Mrk~421 \citep{DiGesu2022}, 1ES~1959+650 \citep{Errando2024}, 1ES~0229+200 \citep{Ehlert2023}, PG~1553+113 \citep{Middei2023}, and PKS~2155-304 \citep{Kouch2024} provided further support for this emerging picture. 

Arguably, the most interesting X-ray polarization behavior has been found in Mrk~421. While the first observation found a similar trend of X-ray higher polarization than that at longer wavelengths, as for Mrk~501, two subsequent observations revealed the first detection of an X-ray polarization angle rotation \citep{DiGesu2023}. Rotations of the polarization angle are often detected at optical wavelengths \cite[e.g.,][]{Marscher2008_nature_MW_EVPA_rot,Blinov2015_robopol_evpa_rot_p1,Blinov2018_robopol_evpa_rot_p2,Liodakis2020}, with long-term rotations associated with gamma-ray flaring \citep{Blinov21}.  But so far, no clear single cause has been identified. Further change in the X-ray polarization angle was detected in the fourth {\it IXPE} observation \citep{Kim2024}, which is possibly associated with the ejection of a new moving radio knot found in contemporaneous VLBA images, although the connection between the X-ray rotation and radio ejection requires further investigation.  This would, however, be in line with the optical polarization behavior of some blazars with lower synchrotron SED peaks \cite[$\nu_{\rm syn}<10^{14}$ Hz,][]{Marscher2008_nature_MW_EVPA_rot,Marscher2010,Liodakis2020} than those of Mrk~421 and Mrk~501, whose SEDs peak at X-ray energies, classifying them as HSP sources \citep[e.g.,][]{Ajello2022}.

Here we present the first dense X-ray polarization monitoring of a blazar, Mrk 421, with {\it IXPE} measurements carried out in four segments during 2023 December 6--22, the last blazar observations during {\it IXPE}'s two-year prime mission. In Section \ref{sec:obs} we describe our X-ray and multi-wavelength observations, in Section \ref{sec:results} we present our results, and in Section \ref{sec:discussion} we discuss our findings.

%%%%%%%%%%%%%%%%%%%%%%%%%%%%%%%%%%%%%%%%%%%%%%%%%%%%%%%%%%%%%%%%%%%%%%%%%%%%
\section{Observations and Data Reduction}
\label{sec:obs}

\begin{deluxetable*}{lcll}
\tablecaption{Summary  of X-ray Observations.\label{tab:xrayobs}}
\tablehead{
\colhead{Observatory} & 
\colhead{Obsid} &
\colhead{Start Date} &
\colhead{End Date}
}
\startdata
IXPE &02008199	&2023-12-06 21:28:11.792 & 2023-12-22 07:44:27.400\\
NuSTAR&60902024002	&2023-12-06 23:26:00	&  2023-12-07 10:31:09\\
NuSTAR&60902024004	&2023-12-11 12:36:00	&	2023-12-12 00:11:09\\
NuSTAR&60902024006	&2023-12-18 06:41:09 & 2023-12-18 18:16:09\\
NuSTAR&60902024008	&2023-12-20 19:31:00	& 2023-12-21  06:41:09	\\
XMM&0902112401 &2023-12-07 00:16:39 &2023-12-07 03:53:19	\\
XMM&0902112501 &2023-12-08 12:53:01 &2023-12-08 18:16:21\\
XMM&0920900201 &2023-12-11 12:52:35  &2023-12-11 18:08:23\\
XMM&0920901301 &2023-12-14 04:14:29 &2023-12-14 10:54:29\\
\enddata

\end{deluxetable*}

\subsection{X-rays}

Mrk 421 was observed by {\it IXPE} in 2023 between December 6 and December 22, as well as by {\it NuSTAR} and {\it XMM-Newton} over multiple epochs during this time period.  The observations used in our analysis are listed in Table \ref{tab:xrayobs}.

\subsubsection{NuSTAR}
NuSTAR observed Mrk 421 four times during the {\it IXPE} campaign, on 2023 Dec 6 for a net focal plane module A/B exposure time (respectively) of 21.29/21.13 ks \footnote{The focal plane module A \& B (FPMA,FPMB respectively) exposure times are written as FPMA/FPMB.} (ObsID: 60902024002), on 2023 Dec 11 for a net exposure time of 21.2/21.1 ks (ObsID: 60902024004), on 2023 Dec 18 for a net exposure time of 17.4/17.4 ks (ObsID: 60902024006), and on 2023 Dec 20 for a net exposure time of 20.6/20.4 ks \arcsec(ObsID:60902024008). The {\it NuSTAR} spectra were extracted using a circular $1$\arcmin\ radius source region and a circular $100$\arcsec\ radius background region. The spectral extraction was accomplished with \texttt{HEASOFT} version 6.33, \texttt{NUSTARDAS} version 2.1.2, and CALDB version 20240229. Each spectrum was grouped to have a minimum of 30 counts per bin. Above 50 keV the background became stronger than the source, so we limit {\it NuSTAR} photon energies to 3.0--50.0 keV in spectral fits which incorporate {\it NuSTAR} data (see Section \ref{sec:IXPEdata}). The instrumental cross-normalization constant for FPMB was set to be 1.04 times the normalization constant for FPMA, as in \citet{madsen2015}.

\subsubsection{XMM-Newton}

{\it XMM-Newton} observed Mrk 421 four times during the {\it IXPE} campaign: on 2023 Dec 7 for a duration of 13.0 ks (Obsid: 0902112401),  on 2023 Dec 8 for a duration of 19.4 ks (Obsid: 0902112501), on 2023 Dec 11 for an exposure time of 20.0 ks (Obsid: 0920900201), and on 2023 Dec 14 for an exposure time of 24.0 ks (Obsid: 092091301).  All observations used the timing data mode and thick filter for the EPIC instrument.  We reprocessed the observations using {\it XMM} SAS release 20230412\_1735-21.0.0 and filtered the light curves for background flares.

MOS1 data were unavailable for analysis. We initially extracted spectra from the full timing window for both MOS2 and PN cameras using standard timing mode threads. To test for photon pile-up, we used the {\it epatplot} tool, and found a $\sim20\%$ excess of double events.  We fit {\it NuSTAR} spectra to an absorbed power law and used {\tt WebPIMMS}\footnote{\href{https://heasarc.gsfc.nasa.gov/cgi-bin/Tools/w3pimms/w3pimms.pl}{https://heasarc.gsfc.nasa.gov/cgi-bin/Tools/w3pimms/w3pimms.pl}} to predict PN pile-up.  The predicted PN count rate was well below pile-up thresholds and predicted pile-up rates were negligible.  As a precaution against pile-up, we used only single events and excised the brightest ‘core’ rows by iteratively masking them from the spectral extraction. We increased the size of the mask by one pixel in each direction per iteration, until the best fit to a simple absorbed power law produced negligible ($<1$\%) change in $n_H$, power law index, and normalization.  Given the low expected pile-up, the actual goodness of the power law model fit is irrelevant for determining the exclusion zone, only the variation in the model approximation in order to determine the row-specific impact of pile-up.
\subsubsection{IXPE}
\label{sec:IXPEdata}
\begin{figure*}
    \centering
    \begin{tabular}{cc}
    \begin{minipage}{0.99\textwidth}
    \includegraphics[width=\textwidth,angle=0]{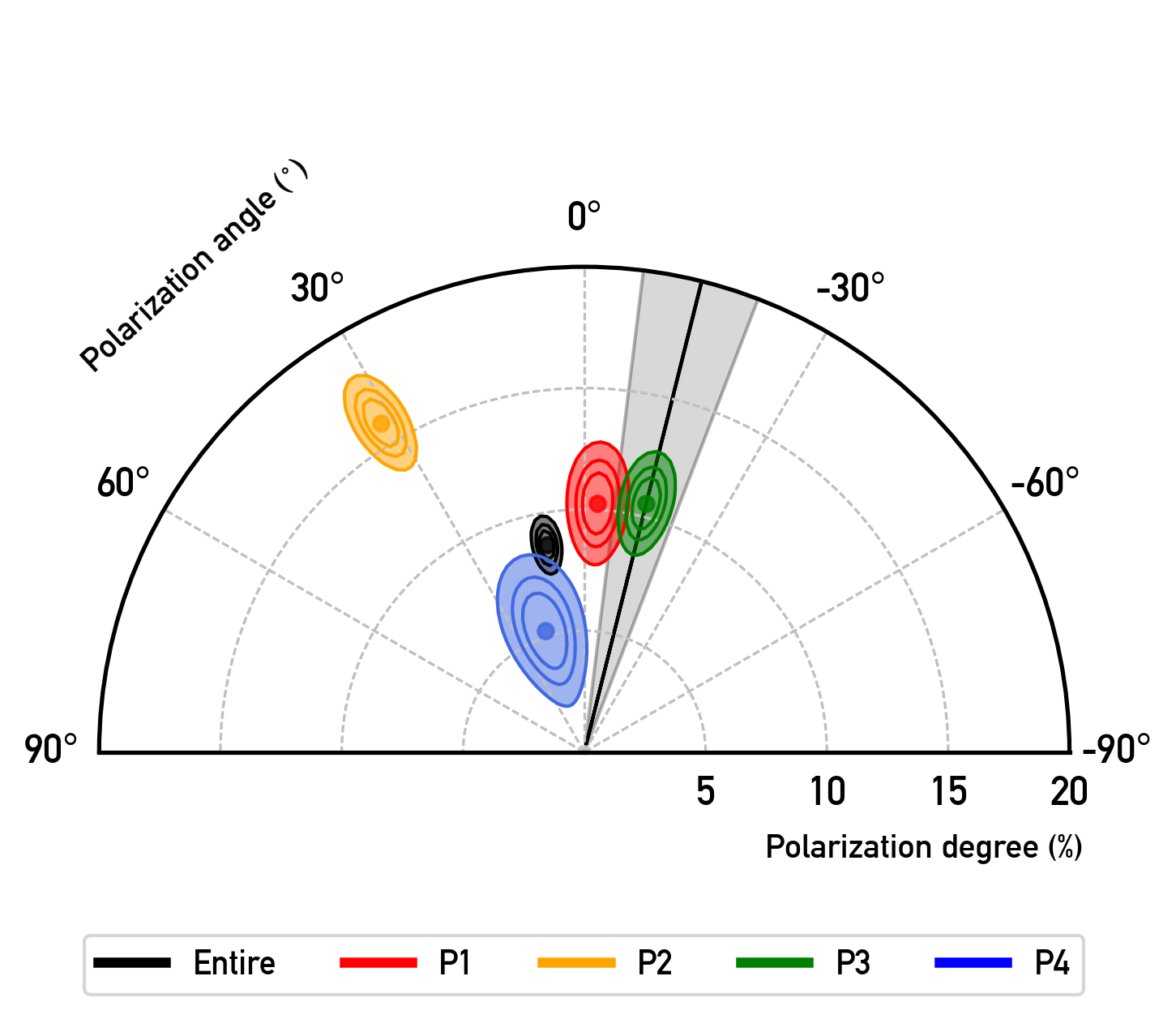}\end{minipage}\\
\begin{minipage}{0.99\textwidth}
    \includegraphics[width=0.24\textwidth]{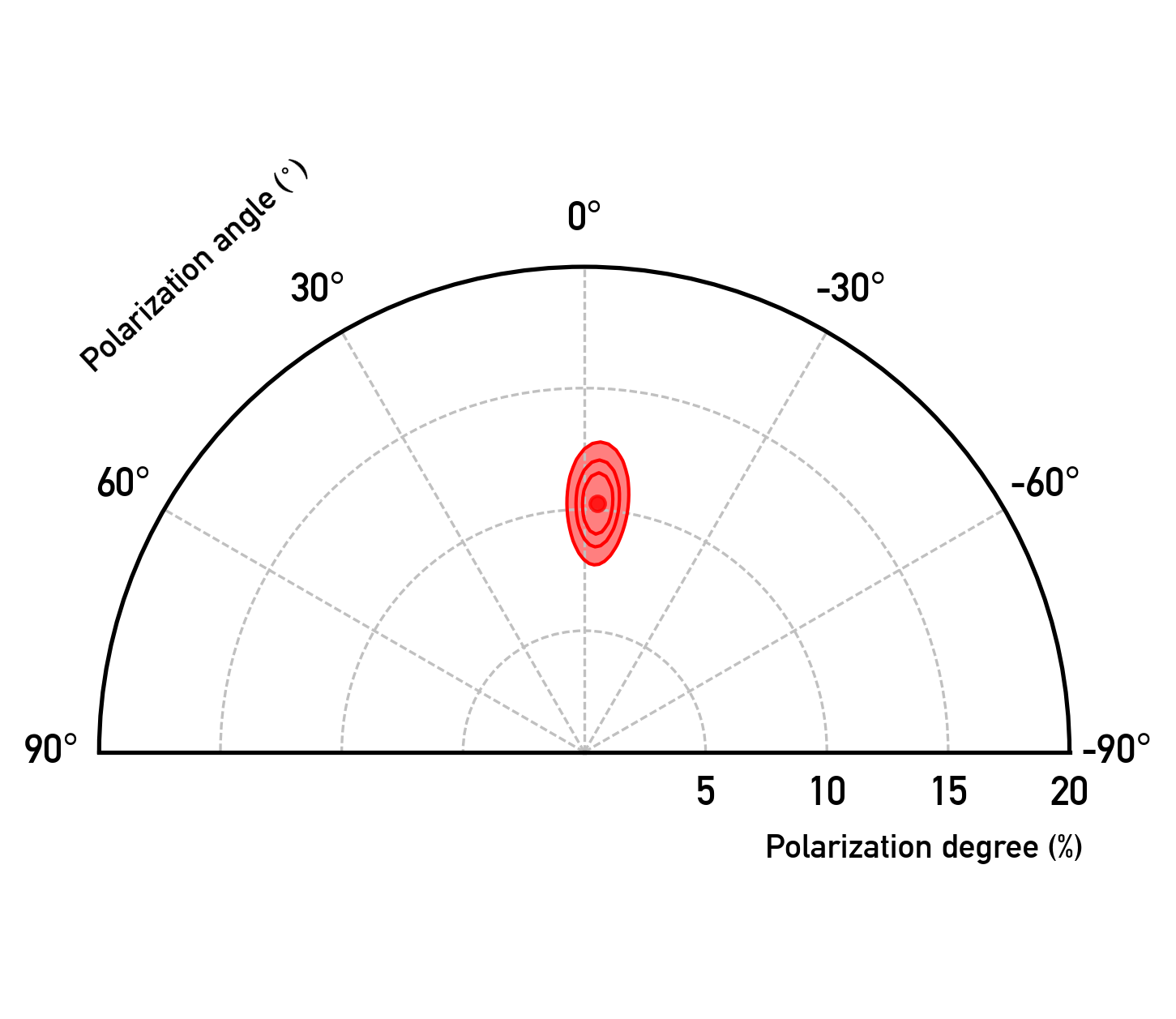} \includegraphics[width=0.24\textwidth]{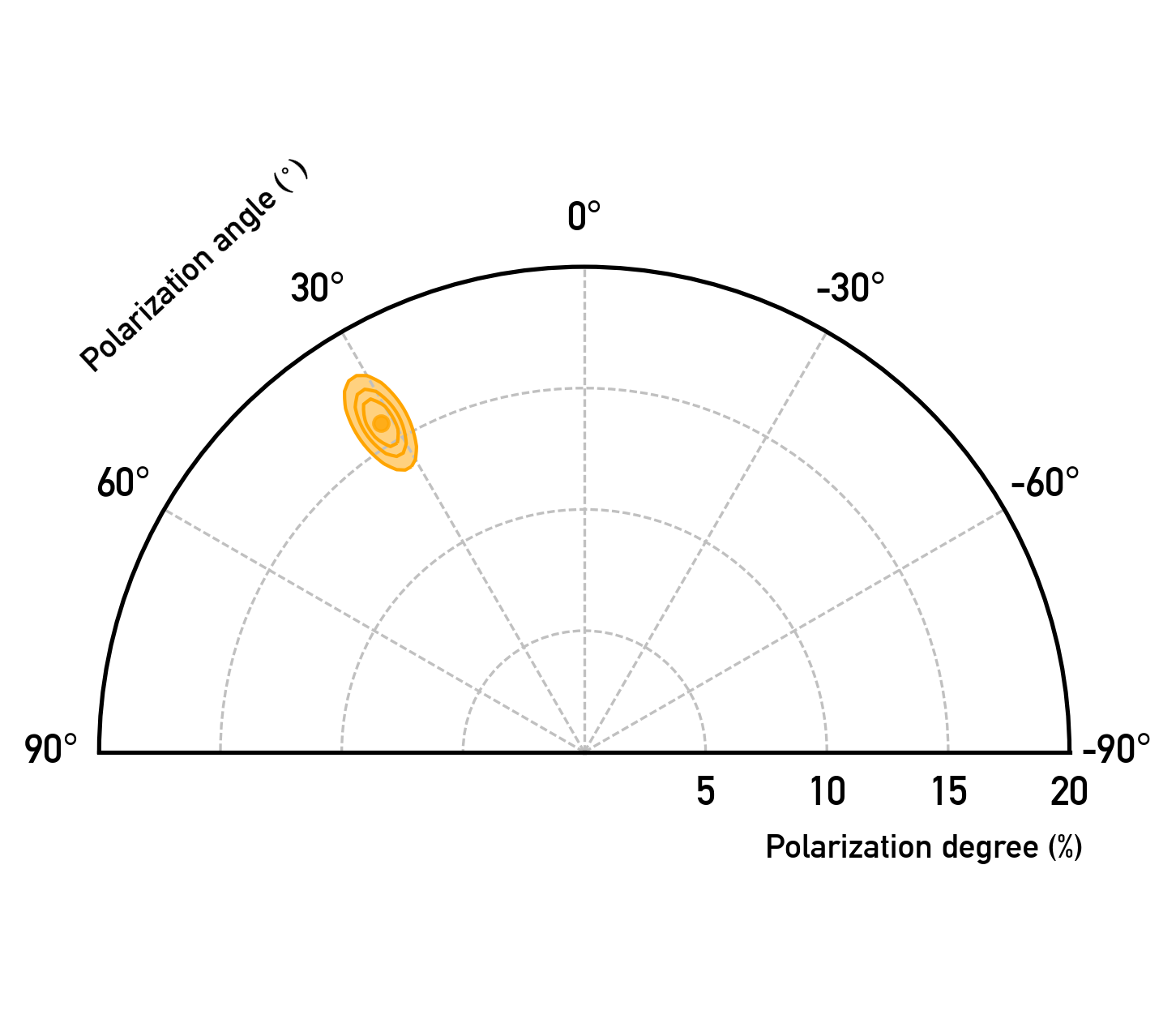}
    \includegraphics[width=0.24\textwidth]{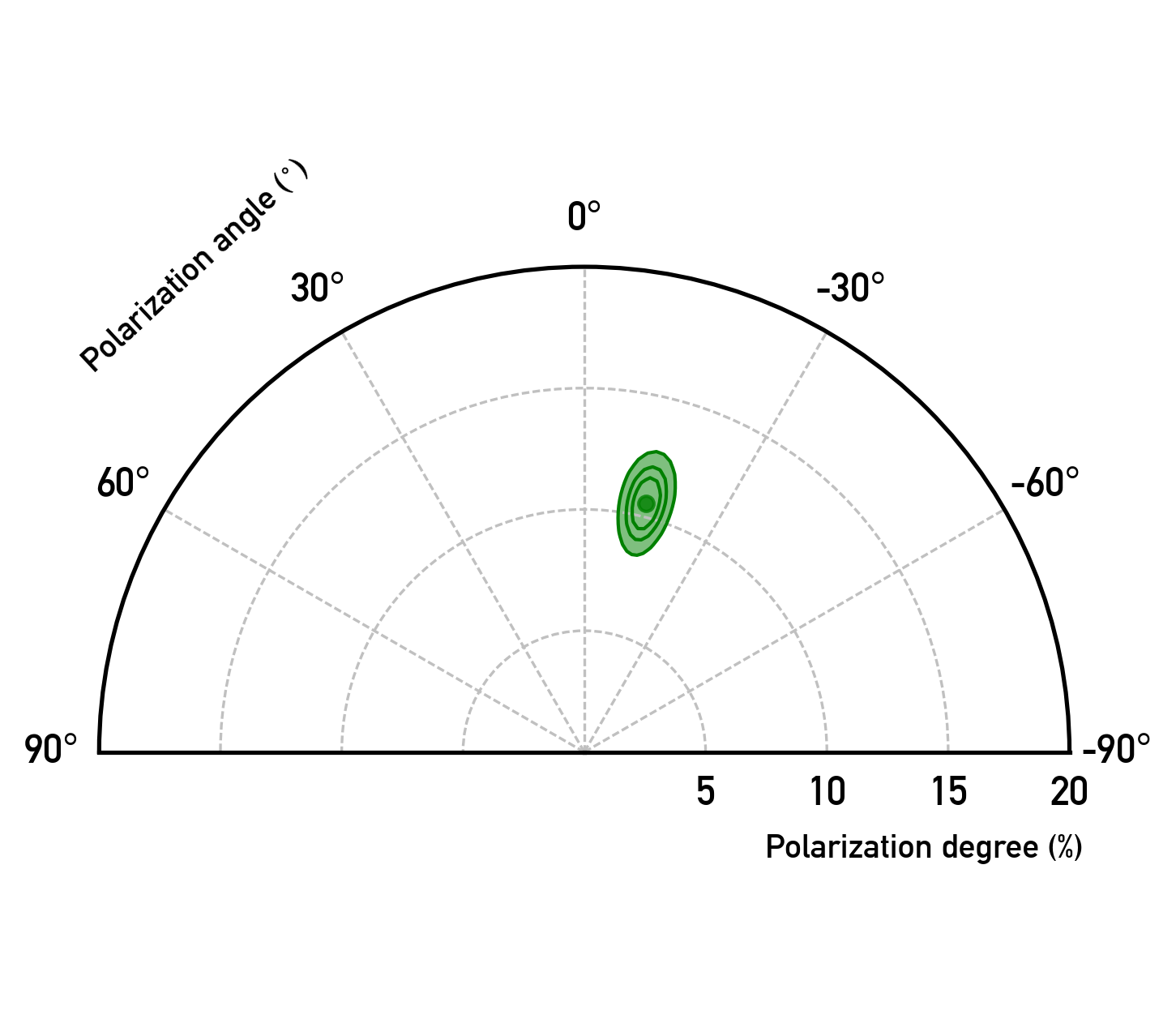}
    \includegraphics[width=0.24\textwidth]{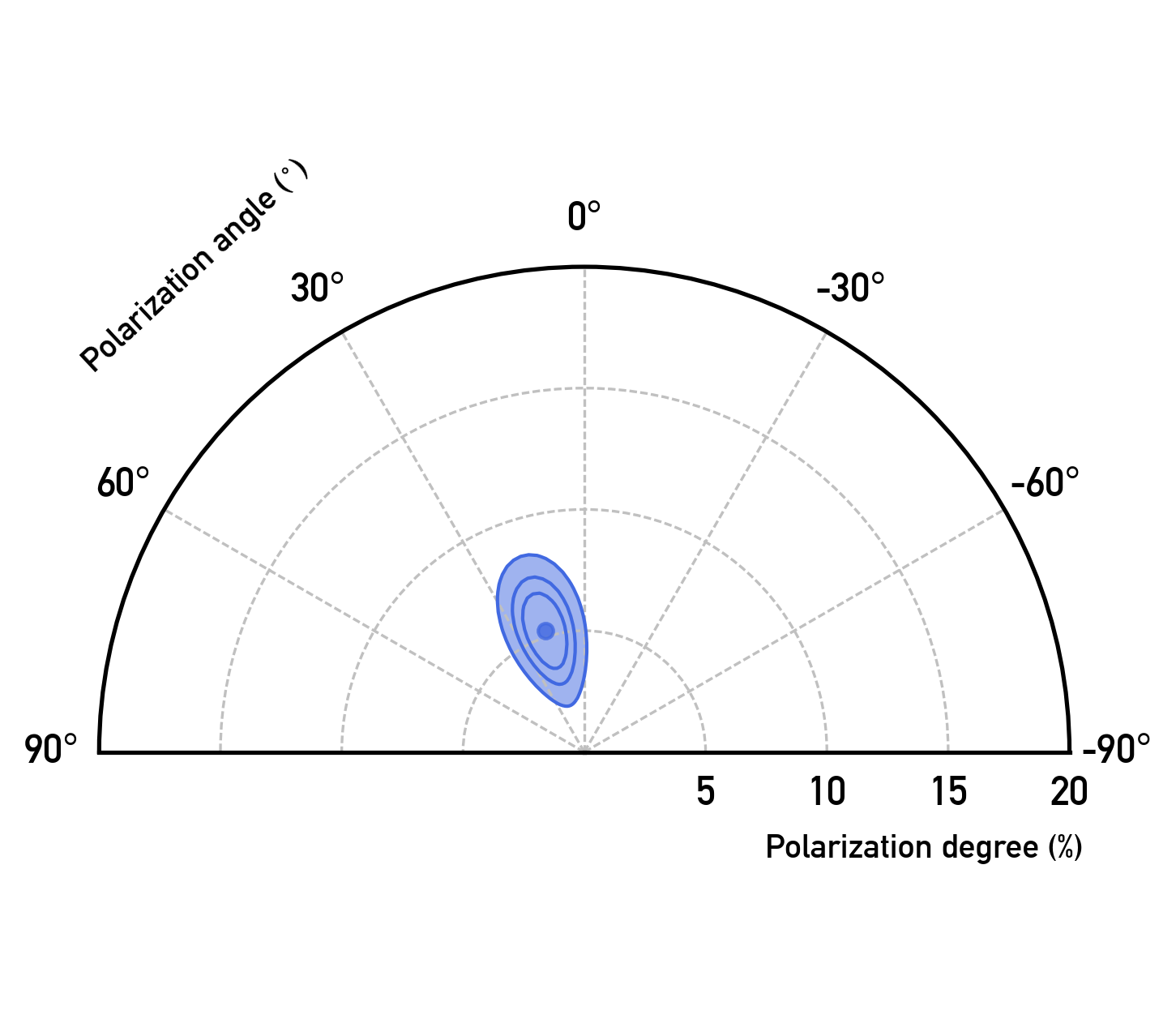}\end{minipage}
    \end{tabular}
    \caption{Time-averaged X-ray polarization contours for each of the four {\it IXPE} epochs for this campaign, labeled P1 (red), P2 (yellow), P3 (green) and P4 (blue), respectively, in order of time, shown together (top) and separately (bottom).  ``Entire'' (black) represents all data for the campaign.  Radial offset from the origin corresponds to polarization degree $\Pi$, and angle corresponds to polarization angle $\psi$  Contours are drawn at 68.27\%, 90.00\%, and 99.00\% confidence levels, according to a $\chi^2$ test with two degrees of freedom. The black line and gray shading in the unified plot (top) represents the position angle of the jet on the sky in degrees, as observed by the VLBA at 43 GHz.}
    \label{plt:polplot}
\end{figure*}

{\it IXPE} observed Mrk 421 during four separate pointings from 2023 Dec 6 to 2023 Dec 22, spanning 15.4 days  (ObsID: 02008199).  The pointings cover Mrk 421 for 927 ks, of which 514 ks is exposure (after accounting for factors such as earth eclipses). The four observation periods are simultaneous with the four {\it NuSTAR} observations. Using the same methods as \cite{Kim2024}, we calculated the polarization angle and degree for each pointing.  These are shown in Figure \ref{plt:polplot}.
The $I$, $Q$, and $U$ spectra were extracted using \texttt{ixpeobssim} \citep{baldini22} version 30.6.3, within a circular source region of $1'$. The background was subtracted from an annular region with inner radius $2'$ and outer radius $5'$.  The CALDB IRF version 13 (date 2023/07/02) in \texttt{ixpeobssim} was used for calibration.  The spectra were weighted according to instrumental response. The $I$ spectrum was grouped with a minimum of 30 counts/bin, while the $Q$ and $U$ spectra were grouped within equal 0.2 keV width bins. The polarization measurements were performed using the \texttt{pcube} algorithm \citep{baldini22} and were background subtracted with a backscale ratio of 1/20. The spectra used in the spectropolarimetric fit were also background subtracted. The energy range 2.0-8.0 keV was used for the entire {\it IXPE} analysis.

\paragraph{Spectropolarimetric Fitting}
For each {\it IXPE} observation, we jointly fit the I, Q, and U spectra together with the corresponding simultaneous {\it NuSTAR} observation. Both {\it XMM } observations coincided with the first {\it IXPE} observation, so we jointly fit them only with the first {\it IXPE} and first {\it NuSTAR} observations.
 
We started with a \texttt{constant*tbabs*polconst* zlogpar} model, where \texttt{constant} is the instrumental cross-normalization constant, \texttt{tbabs} represents Galactic absorption, and \texttt{zlogpar} is a redshifted log-parabolic model similar to what has been used in the past to fit Mrk 421's X-ray spectrum \citep{Massaro2004,DiGesu2022}. The spectral distribution is expressed in the form 
\begin{equation}
N(E)=K(E(1+z)/E_{p})^{(\alpha-\beta log(E(1+z)/E_{p}))}.
\end{equation}
In this equation, $E_{p}$ is the pivot energy, $\alpha$ is the spectral slope at the pivot energy, $\beta$ is a spectral curvature term, and $K$ is a normalization constant. The units of $K$ are $\mathrm{photons/(cm^{2}\;s\;keV)}$. We fix the pivot energy to 5 keV \citep[e.g.,][]{Balokovic2016,DiGesu2022,Middei2022}. We attempted to add a \texttt{zphabs} model to take into account absorption at the source, but it did not improve the fit, so it was dropped. We therefore fit all four {\it IXPE} observations with a \texttt{constant*tbabs*polconst*zlogpar} model.

The results of the fits are tabulated in Table \ref{tab:spectropol}. The columns sort the results by the corresponding {\it IXPE} observation, while the rows represent the time-averaged PD, the time-averaged PA, $\alpha$, $\beta$, the rest frame 2-10 keV flux, and the reduced $\chi^2$ for each observation. The reduced $\chi^2$ values of the fits range between 1.08 and 1.34, suggesting that while our model may be a reasonable first approximation, there is a need for additional model complexity.  The {\it XMM} spectra in particular have notably strong residuals at some energies. The PD varies from $6\pm2\%$ to $20\pm1\%$ between observations, while the PA ranges from $-14\pm3\degr$ to $+32\pm2\degr$. These spectra and fits are plotted in Appendix \ref{sec:appA}.

\begin{deluxetable}{lcccc}
\setlength{\tabcolsep}{2pt}
\tablecaption{Results of spectropolarimetric fitting for \texttt{constant*tbabs*polconst*zlogpar} model. We note that the values for $\alpha$ and $\beta$ are those from NuSTAR.\label{tab:spectropol}}
\tablehead{
\colhead{} & 
\colhead{Int 1} &
\colhead{Int 2} &
\colhead{Int 3} &
\colhead{Int 4}
}
\startdata
PD(\%) & $12\pm2$ & {$19\pm1$} & {$12\pm1$} & {$6\pm2$}\\
PA(\degr)\tablenotemark{a} & $-2.6\pm3.9$ & {$32\pm2.2$} & {$-14\pm3.2$} & {$19\pm9.7$}\\
$\alpha$ & $2.8\pm0.01$ & {$2.8\pm0.01$} & {$2.9\pm0.01$} & {$2.9\pm0.01$}\\
%$\beta$ & $(1.7\pm0.05)\times10^{-1}$ & {$(3.7\pm0.18)\times10^{-1}$} & {$(2.3\pm0.18)\times10^{-1}$} & {$0.33\pm0.02$}\\
$\beta$ & $0.170\pm0.05$ & {$0.37\pm0.02$} & {$0.23\pm0.02$} & {$0.33\pm0.02$}\\Flux\tablenotemark{b} & {$1.85^{+0.01}_{-0.01}$}& {$2.27^{+0.03}_{-0.03}$} & {$2.90^{+0.01}_{-0.01}$} & {$2.64^{+0.03}_{-0.03}$}\\
$\chi^{2}/d.o.f.$ & {2122/1583} & {1416/1245} & {1355/1253} & {1438/1215}
\enddata
\tablenotetext{a}{Measured counterclockwise from west to east, $-90\degr$ to $+90\degr$}
\tablenotetext{b}{The 2-8 keV observed flux in units of $10^{-10}\;\mathrm{erg\;cm^{-2}\;s^{-1}}$}
\end{deluxetable}

\paragraph{Time variability}

\begin{figure*}
    \centering
    \includegraphics[width=\textwidth]{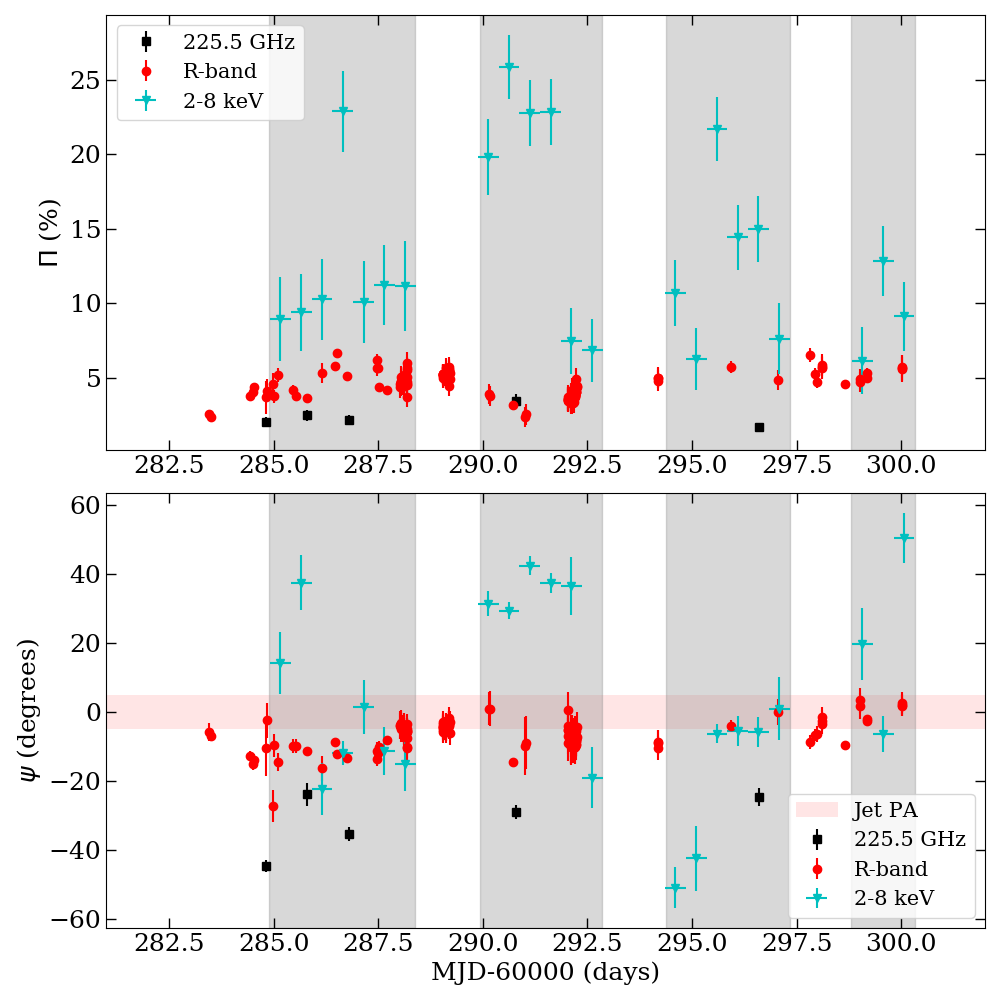}
    \caption{Time-resolved multi-wavelength polarization observations. The top panel shows the polarization degree and the bottom panel shows the polarization angle, relative to the jet. The gray shaded areas in both panels mark the duration of the {\it IXPE} exposures, and the red shaded area in the bottom panel indicates the direction of the jet on the plane of the sky. We only show bins where the X-ray polarization degree is detected at the $>3\sigma$ level.}
    \label{plt:resolved_xpol}
\end{figure*}

We investigated polarization variability over time by applying the $\chi^2$ test of the constant model on different time scales of binning over the entire observation period, following \cite{Kim2024}. This test examines the statistical significance of the variability by estimating the null hypothesis probability ($P_{\text{Null}}$) of each constant model fitting on the normalized $q$ and $u$ Stokes parameters, while taking into account the uncertainty of each measurement. We divided the entire observation period into 10 to 35 sub-periods, corresponding to $\sim$ 130 ks to $\sim$ 40 ks, respectively. As a result, we obtained less than 1\% for the null hypothesis probability of the constant model for all cases of time binning of $q$ and $u$ Stokes parameters. The highest null hypothesis probability for $q$ was $2.2\times10^{-8}$, which corresponds to an out-of-range value of $\sim$ 5.5$\sigma$ ($\sim5.0\sigma$ for $\ge1$ bin of 31), and we measured zero probability for the $u$ Stokes parameter. Hence, we found evidence of statistically significant polarization variability over time from this observing period. Figure \ref{plt:resolved_xpol} shows the polarization measurements split in 31 bins (of size $\sim43$ ks per bin), which provided the largest number of statistically significant polarization variations.

\begin{figure*}
    \centering
    \includegraphics[width=\textwidth]{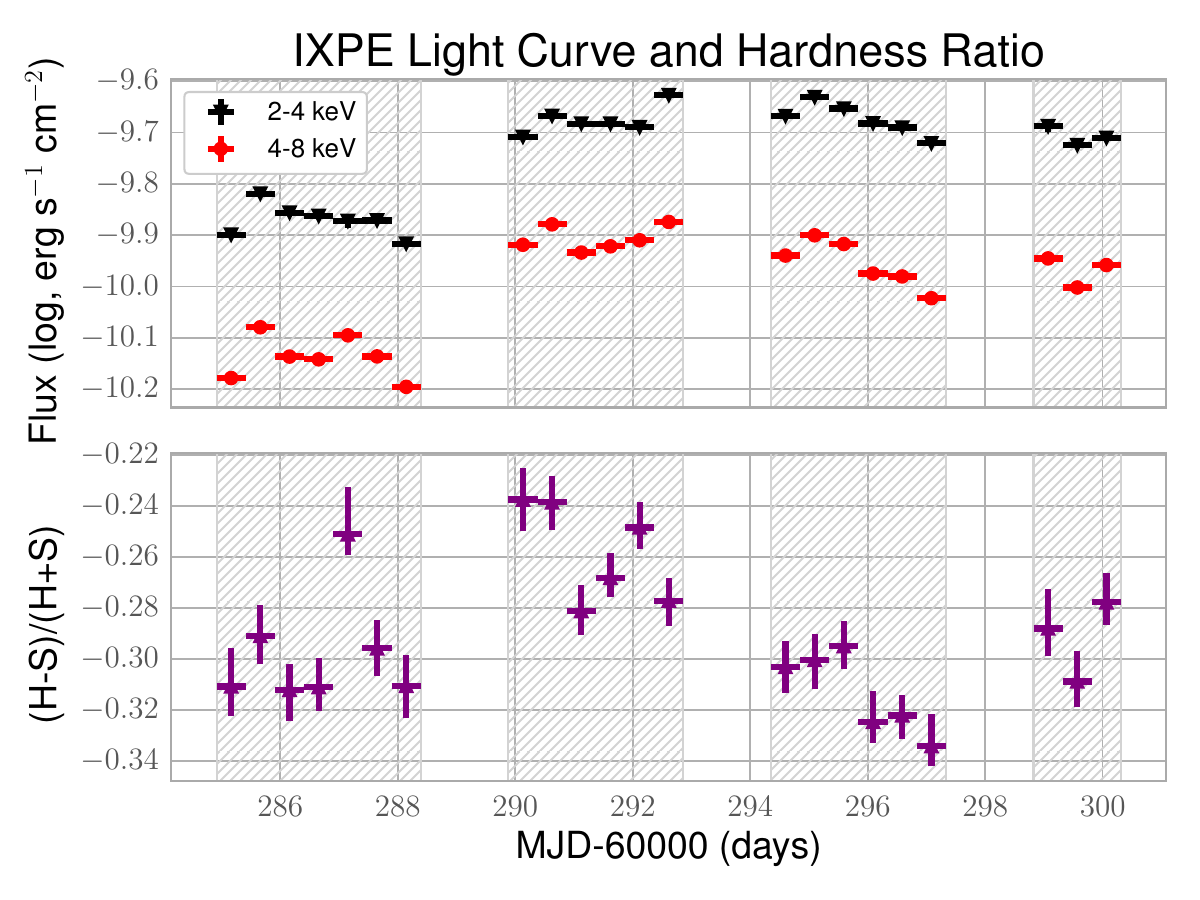}
    \caption{Top: light curves of the log$_{10}$ {\it IXPE} flux for soft (2-4 keV, black triangles) and hard (4-8 keV, red dots) bands.  Bottom: hardness ratios calculated from the soft and hard flux.
    \label{fig:hardness}}
\end{figure*}

In order to compare polarimetric variability with changes in the X-ray flux and spectral shape, we calculate the hard (H; 4-8 keV) and soft (S; 2-4 keV) fluxes, as well as a hardness ratio HR as a proxy for spectral shape, using the same  $\sim43$\,ks binning scheme as with time-resolved polarimetry.  We define the hardness ratio $HR=(H-S)/(H+S)$, where H is the 4-8 keV flux and S is the 2-4 keV flux.  The variation in these quantities is shown in Figure \ref{fig:hardness}.  We note a flare in both $H$ and $S$ by a factor of $\sim2$ at $MJD\sim60289$, after which the flux gradually decays over the next $\gtrsim10$ days, with significant variability on $\sim43$\,ks scales but remains elevated.  The large flare is initially harder when brighter (as with \citealt{DiGesu2023}, but this trend does not uniformly continue for the rest of the campaign.  The softest points (at $MJD\sim60297$) occur during this elevated, decaying state, after which the hardness (but not flux) becomes comparable to pre-flare conditions.

\subsection{Optical observations}

Mrk~421 was observed by several telescopes in BVRI bands during all four segments of the {\it IXPE} observations. These facilities included the Kanata telescope using the Hiroshima Optical and Near-InfraRed camera (HONIR, \citealp{kawabata_new_1999,akitaya_honir_2014}), Liverpool Telescope using the Multicolour OPTimised Optical Polarimeter \cite[MOPTOP,][]{Jermak2016,Shrestha2020}, LX-200 telescope of St. Petersburg State University \citep{lx200}, Nordic Optical Telescope (ALFOSC, analysis described in \citealt{Hovatta2016} and \citealt{Nilsson2018}), Boston University's 1.8 m Perkins Telescope (with the PRISM camera, \citealt{BUPerkins}), Sierra Nevada Observatory \cite[DIPOL-1,][]{juan_escudero:2023,Otero-Santos2024,Escudero2024}, Skinakas Observatory \cite[RoboPol,][]{Ramaprakash2019}, and the T-60 telescope at the Haleakala observatory \citep{Piirola1973,Berdyugin2018,Berdyugin2019,Piirola2021}. We corrected for dilution of the polarization by unpolarized host-galaxy light in the R-band by estimating the flux density of the host within a given aperture and subtracting from the total flux density following \cite{Nilsson2007} and \cite{Hovatta2016}. Details of the data reduction and analysis of the different participating telescopes can be found in \cite{Liodakis2022nature}, \cite{Peirson2023}, and \cite{Kouch2024}. The optical flux and polarization measurements are shown along with the {\it IXPE} results in Appendix \ref{sec:appB}. The points labeled as R-band with a ``$^\dagger$" correspond to R-band observations that we were unable to correct for the host-galaxy contribution, but are provided here for relative comparison. 

\subsection{Radio observations}
Radio observations covered frequencies from 4.8~GHz to 225.5~GHz using the Effelsberg 100-m radio telescope \cite{Kraus2003}, the Korean VLBI Network (KVN), and the Submillimeter Array (SMA)  polarimeter \citep{Marrone2008}. The Effelsberg 4.8~GHz and 14.2~GHz observations are part of the  program Monitoring the Stokes Q, U, I and V Emission of AGN jets in Radio (QUIVER, \citealp{Myserlis2018}). The KVN provided 25~GHz observations using the Yonsei and Tamna antennas in single-dish mode \citep{Kang2015}, and the SMA observations were performed as part of the SMA Monitoring of AGNs with POLarization (SMAPOL, Myserlis et al., 2024 \textit{in preparation}) program at 225.5~GHz. Data from all  radio observations are shown in Appendix \ref{sec:appB}.
\begin{figure*}
    \centering
\includegraphics[width=\textwidth]{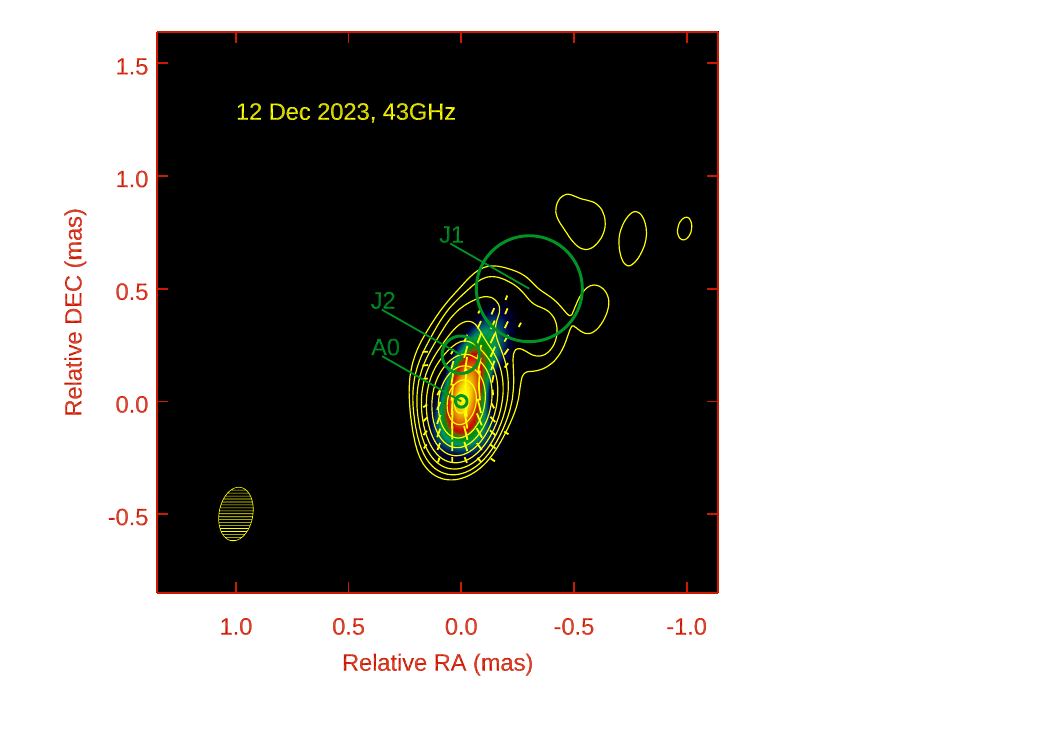}
    \caption{Total (contours) and polarized (color scale) intensity images of Mrk~421 at 43 GHz, convolved with an elliptical Gaussian beam of dimensions 0.24$\times$0.15 mas$^2$ with major axis along $PA=-10^\circ$. The global total intensity peak is 321 mJy/beam and the global polarized intensity peak is 9.6 mJy/beam. Yellow line segments within the image indicate the direction of polarization, while the length of each segment is proportional to the polarized intensity. Green circles designate the core, $A0$, and jet components, $J1$ and $J2$, according to model fitting to the uv data.}
    \label{fig:vlbaimage}
\end{figure*}

\begin{deluxetable*}{lrrrrrr}
\tablecaption{Parameters of Jet Components at 43 GHz}\label{tab:vlbaparam}
\tablehead{\colhead{Knot}&\colhead{$S$,mJy}&\colhead{$R$,mas}&\colhead{$PA_{\textit jet}$,deg}&
\colhead{Size,mas}&\colhead{$PD_{\textit 43GHz}$,\%}&\colhead{$EVPA_{\textit 43GHz}$,deg}\\
(1)&(2)&(3)&(4)&(5)&(6)&(7)
}
\startdata
$A0$&335$\pm$15&0.0&\nodata&0.05$\pm$0.02&3.0$\pm$0.4&2$\pm$4\\
$J2$&37$\pm$8&0.21$\pm$0.05&1$\pm$5&0.17$\pm$0.03&7.0$\pm$1.5&$-$25$\pm$6\\
$J1$&25$\pm$10&0.52$\pm$0.09&$-$38$\pm$7&0.4
7$\pm$0.10&18.0$\pm$7.5&$-$12$\pm$10\\
\enddata
\vspace{0.05in}
\noindent Column identification: 1 - name of component; 2 - flux density, $S$, milliJanskys; 3 - distance from the core, $R$, milliarcseconds (mas), 4 - position angle of the knot with respect to the core, $PA_{\text jet}$, degrees; 5 - FWHM size of the component, mas; 6 - degree of polarization, percentage; 7 - position angle of polarization, degrees.   All uncertainties are $1\sigma$. 
\end{deluxetable*}
In the beginning of the second {\it IXPE} pointing, 2023 December 12 (RJD=289, where RJD=JD$-$2460000.0),
we obtained total and polarized intensity images of Mrk~421 with the Very Long Baseline Array (VLBA) at 43~GHz under the Boston University BEAM-ME project \footnote{www.bu.edu/blazars/BEAM-ME.html}. The data reduction was performed using the Astronomical Image Process System (AIPS) and Difmap software packages, as described in \citet{Jorstad2017}. 
Figure~\ref{fig:vlbaimage} displays the total intensity image with the polarized intensity superposed, along with line segments indicating the EVPA.
We have modeled the total intensity image by components with circular Gaussian brightness distributions. We measured the Stokes Q and U flux densities of these components (knots) and derived their PD and EVPA at 43 GHz. Table~\ref{tab:vlbaparam} lists the total intensity and polarization parameters of the three brightest components ($A0$, $J1$, and $J2$). This model fits the uv data very well, with a reduced $\chi^2$ of 0.86. 
The location and size of each component is marked in Figure~\ref{fig:vlbaimage} by green circles. Knot $A0$, located at the southern end of the jet, represents the VLBI core of the jet, while the jet component nearest to the core, $J2$, defines the innermost jet direction as $PA_{\textit jet}=1^\circ\pm5^\circ$. This is, within the uncertainties, parallel to the parsec-scale jet direction of Mrk~421, $PA_{\textit jet}=-14^\circ\pm14^\circ$, estimated based on 92 epochs of observations from 2007 to 2018 \citep{Weaver2022}.  

%%%%%%%%%%%%%%%%%%%%%%%%%%%%%%%%%%%%%%%%%%%%%%%%%%%%%%%%%%%%%%%%%%%%%%%%%%%%%
\section{Results}\label{sec:results}

To probe short-term variability, we divide the {\it IXPE} observations into 31 bins as discussed above. Figure \ref{plt:resolved_xpol} shows the significantly detected X-ray polarization bins. Only one bin does not have a significant detection. We find rapid variability in both the polarization degree and angle. Between the second and third observations we also detect a significant rotation of the polarization angle by $\sim90^\circ$. This is significantly less than the previously detected rotation in Mrk 421 of about 400$^\circ$ \citep{DiGesu2023}. The direction of the polarization vector seems to vary about the jet axis, which is typically of HSP blazars (\citealt{Marscher2021}, \citealt{Chen24}, Lisalda et al \textit{submitted}).
%%\textit{\color{green} add lisalda, chien-tings paper}. 
We discuss the origin of the rotation below.

On the other hand, at radio wavelengths the degree of polarization of Mrk~421 is low (1-3\%), with a median of $\sim2\%$. In the past, the mm-radio polarization angle has been fairly stable along $-29^\circ\pm8^\circ$, consistent with the jet axis on the sky over long time-scales, $-14^\circ\pm14^\circ$ (see above). However, the direction of both the jet and the EVPA at 43 GHz in the innermost regions was
$1\pm5^\circ$ on 12 December 2023 (MJD=60290) during the {\it IXPE} campaign (see Fig.\ \ref{fig:vlbaimage}), identical with the optical EVPA. The EVPA of SMAPOL and QUIVER data exhibits a relatively stable frequency dependence across cm and mm wavelengths , suggestive of a Faraday screen with rotation measure $\rm RM\sim-250~rad/m^2$ \citep{Manchester72}.

We find a similar behavior at optical wavelengths: there are no large variations in either flux, polarization degree, or polarization angle. The polarization degree smoothly changes from 2 to 6\% with a median of about 4.5\%. The polarization angle is roughly stable and  along $-7^\circ$, a bit offset from the jet axis. Compared to previous observations contemporaneous with {\it IXPE}, the source maintains consistent behavior at radio and optical wavelengths \citep{Kim2024}. 

The 31-bin light curve of the {\it IXPE} soft and hard flux is matched to the polarimetric binning and within any given epoch is consistent with increased spectral hardness at high flux, with the exception of epoch 2 ($290<\rm{MJD}-60000<293$), where there is no obvious correlation over a timescale of $\sim$days.  After the first epoch ($284<\rm{MJD}-60000<289$), the flux increases by a factor of $\sim2$ and remains comparably high for the remainder of the campaign.  This flare corresponds to an increase in hardness, $\sim6\sigma$ above the mean value for the {\it IXPE} campaign ($HR\sim0.295$), as well as an increase in polarization degree from $\sim10\%$ to $\sim25\%$ which is sustained over $\sim2$ days.  The most extreme deviations of polarization angle from the jet value seem to be associated with either very hard ($\rm{MJD}-60000\sim290$) or very soft ($\rm{MJD}-60000\sim296$) spectra.  Epochs 2 ($290<\rm{MJD}-60000<293$) and 3 ($294<\rm{MJD}-60000<298$) show a trend of decreasing HR, to below pre-flare levels, and the $\sim90^\circ$ rotation in X-ray polarization angle occurs during this trend, such that the softest HR values are associated with a return to the jet angle, a moderate degree of polarization ($\sim7-15\%$), and rapid variability ($\lesssim$day) of the polarization degree.

\section{Discussion \& Conclusions}
\label{sec:discussion}

We have presented the first dense monitoring of the X-ray polarization of a blazar with {\it IXPE}. We find large variations in the polarization degree and a rotation in the polarization angle. The origin of these rotations requires us to consider models varying from shocks moving along a helical magnetic field down the jet \citep{Marscher2008_nature_MW_EVPA_rot}, bent jets \citep{Abdo2010}, turbulence \citep{Marscher2014},  shock-shock interactions \citep{Liodakis2020}, kink instabilities \citep{Zhang2017}, and magnetic reconnection \citep{Hosking2020,Zhang2020}, among others. \cite{Kiehlmann2017} concluded that relatively small ($\sim90^\circ$) rotations are likely to result from random walks of the magnetic field due to turbulence or emission originating in multiple zones. They also found that larger ($>180^\circ$) rotations are often associated with $\gamma$-ray activity \citep{Marscher2010,Blinov2015_robopol_evpa_rot_p1,Blinov2018_robopol_evpa_rot_p2}, particularly for long-term optical rotation \citep{Blinov21}.

We test the random-walk scenario following \cite{Kiehlmann2017} and \cite{Kiehlmann2021}, in a similar manner as in our previous analysis of the X-ray EVPA rotation detected in Mrk~421 \citep{DiGesu2023}. We generate a large number of simulated polarization light curves by exploring the parameter space of number of cells and number of cells that change per day ($N_{\rm cell}$, $N_{\rm Var}$, respectively). We then identify the simulations that produce similar polarization properties as those observed.

We apply the test to both the entire {\it IXPE} dataset, as well as to only the period of rotation between the second and third observational segments. 
For the former case, we use the median and inter-quantile range of the polarization degree and the inter-quantile range of the polarization angle. We require that the simulations produce values within 10\% of those observed for the individual quantities. We are able to achieve a $>50\%$ success rate between simulations and observations, with the median polarization degree succeeding at a higher rate than the inter-quantile range of the polarization angle. However, when we require all conditions to be satisfied at the same time, the maximum success rate is only 1\%. This is because the observed quantities are better reproduced from different regions of the $N_{\rm cell}$, $N_{\rm Var}$ parameter space, with little overlap. This is consistent with the findings of the larger statistical study of the optical RoboPol sample \citep{Kiehlmann2017}. For segments 2-3, we again require that the median and inter-quantile range of the polarization degree for the simulations be within 10\% of that observed, and that the simulations produce a rotation of the polarization angle that has an equal or larger amplitude. Figures showing the success rate of the simulations for the $N_{\rm cell}$, $N_{\rm Var}$ parameter space are presented in Appendix \ref{sec:appC}. For the polarization degree, we find similar results as in the entire dataset. For the polarization angle, we find that more than 80\% of the simulations are able to produce a rotation that has an amplitude that is equal to, or larger than, the observed value. However, similar to the entire dataset, when we demand that all conditions are met, we can only achieve a success rate of $\sim2\%$. Based on both tests, we conclude that a simple, pure random walk model has only a low probability to reproduce the observed behavior.

Although at least some level of turbulence appears to be present, evident by the lower than theoretically expected polarization degree (see below), turbulence is unlikely to be the primary cause of the EVPA rotation. The observed behavior is likely more akin to a scenario where turbulent plasma passes through a more ordered magnetic field region, such as a shock \citep{Marscher2014}.

During the period of the highly variable X-ray polarization, the radio and optical polarization is at most mildly variable. The X-ray polarization degree reaches a maximum of 25\%, 8$\times$ higher than the simultaneous optical polarization. Throughout our campaign, the X-rays remain consistently more polarized than the radio and optical emission, although there are time bins within which the ratio of X-ray to optical polarization is close to unity. This, along with the time variability of the polarization, suggests that even in the X-ray emission region there is a significant turbulent component.  In general, a role for turbulence is suggested by the fact that we see only 25\% polarization, compared to $\sim$70-75\% expected for synchrotron radiation in a perfectly ordered magnetic field.  Some steady jet models would permit such a low polarization \citep[e.g.,][]{Bolis24}, but would not explain the erratic variability.

In general, the most extreme variability of the X-ray polarization degree during this campaign is observed to occur on shorter timescales ($\sim12$-48 hours over the entire campaign) than for the most extreme flux variability.  The $\rm{MJD}-60000\sim290$ flare is associated with some of the strongest sustained X-ray  polarization observed in Mrk 421 (typically $\lesssim15\%$; \citealt{DiGesu2022}, \citealt{DiGesu2023}, \citealt{Kim2024}), and afterwards the X-ray flux evolves little within a $\pm12\%$ range over $>10$ days.  In comparison to flux variability, polarization variability timescales are shorter by a factor of $>5$ and possibly much more. This is due in part to the vector nature of polarization: for example, $N_{\rm cell}$ regions of equal flux but randomly directed polarizations could maintain a steady total flux, while the polarization degree varies about a mean value of $\langle\Pi\rangle\sim75N_{\rm cell}^{-1/2}\%$ with a standard deviation of $0.5\langle\Pi\rangle$ \citep{Marscher2021}. The polarized emission could also occupy a region that subtends only a fraction of the X-ray emitting region.  The $90^\circ$ rotation occurs over an intermediate timescale ($\sim5$ days).  Harder-when-brighter behavior has previously been observed during June 2022 {\it IXPE} observations of polarization angle rotation in Mrk 421. This is consistent with shock acceleration, which may explain increased polarization in such a case \citep{DiGesu2023}.

The significantly higher polarization and variability strongly points to the X-ray emission originating in a smaller region closer to the acceleration site, where the magnetic field is less disordered, than is the case for lower frequencies. Variations of the polarization angle about the jet axis give further evidence for particle acceleration in shocks \citep{Liodakis2022nature}. The evidence therefore supports the persistent emerging picture of energy-stratified shock acceleration in the jets of HSP blazars \citep[e.g.,][]{Kim2024b}.

%XXX
%\par

%%%%%%%%%%%%%%%%%%%%%%%%%%%%%%%%%%%%%%%%%%%%%%%%%%%%%%%%%%%%%%%%%%%%%%%%
%\section{Conclusions}

%XXX

%\bigskip

%Support for this work was provided by the National Aeronautics and Space Administration through XXX.  We thank XXX. The scientific results reported in this article are based on observations made by (1) the \textit{IXPE X-ray Observatory}, (2) the \textit{NuSTAR X-ray Observatory}, and (3) {\it Neil Gehrels Swift Observatory}. 
\acknowledgements
The Imaging X-ray Polarimetry Explorer (IXPE) is a joint US and Italian mission.  The US contribution is supported by the National Aeronautics and Space Administration (NASA) and led and managed by its Marshall Space Flight Center (MSFC), with industry partner Ball Aerospace (now, BAE Systems).  The Italian contribution is supported by the Italian Space Agency (Agenzia Spaziale Italiana, ASI) through contract ASI-OHBI-2022-13-I.0, agreements ASI-INAF-2022-19-HH.0 and ASI-INFN-2017.13-H0, and its Space Science Data Center (SSDC) with agreements ASI-INAF-2022-14-HH.0 and ASI-INFN 2021-43-HH.0, and by the Istituto Nazionale di Astrofisica (INAF) and the Istituto Nazionale di Fisica Nucleare (INFN) in Italy.  This research used data products provided by the {\it IXPE} Team (MSFC, SSDC, INAF, and INFN) and distributed with additional software tools by the High-Energy Astrophysics Science Archive Research Center (HEASARC), at NASA Goddard Space Flight Center (GSFC).

I.L was supported by the NASA Postdoctoral Program at the Marshall Space Flight Center, administered by Oak Ridge Associated Universities under contract with NASA. The IAA-CSIC co-authors acknowledge financial support from the Spanish "Ministerio de Ciencia e Innovaci\'{o}n" (MCIN/AEI/ 10.13039/501100011033) through the Center of Excellence Severo Ochoa award for the Instituto de Astrof\'{i}sica de Andaluc\'{i}a-CSIC (CEX2021-001131-S), and through grants PID2019-107847RB-C44 and PID2022-139117NB-C44. The Submillimeter Array is a joint project between the Smithsonian Astrophysical Observatory and the Academia Sinica Institute of Astronomy and Astrophysics and is funded by the Smithsonian Institution and the Academia Sinica. Maunakea, the location of the SMA, is a culturally important site for the indigenous Hawaiian people; we are privileged to study the cosmos from its summit. E.L. was supported by Academy of Finland projects 317636 and 320045. 

The research at Boston University was supported in part by National Science Foundation grant AST-2108622, NASA Fermi Guest Investigator grant 80NSSC23K1507,
NASA {\it NuSTAR} Guest Investigator grant 80NSSC24K0565, and NASA Swift Guest Investigator grant 80NSSC23K1145. The Perkins Telescope Observatory, located in Flagstaff, AZ, USA, is owned and operated by Boston University. This work was supported by NSF grant AST-2109127. We acknowledge the use of public data from the Swift data archive. Based on observations obtained with {\it XMM-Newton}, an ESA science mission with instruments and contributions directly funded by ESA Member States and NASA. Partly based on observations with the 100-m telescope of the MPIfR (Max-Planck-Institut f\"ur Radioastronomie) at Effelsberg. 

Observations with the 100-m radio telescope at Effelsberg have received funding from the European Union’s Horizon 2020 research and innovation programme under grant agreement No 101004719 (ORP). S. Kang, S.-S. Lee, W. Y. Cheong, S.-H. Kim, and H.-W. Jeong were supported by the National Research Foundation of Korea (NRF) grant funded by the Korea government (MIST) (2020R1A2C2009003). The KVN is a facility operated by the Korea Astronomy and Space Science Institute. The KVN operations are supported by KREONET (Korea Research Environment Open NETwork) which is managed and operated by KISTI (Korea Institute of Science and Technology Information). 

This work was supported by JST, the establishment of university fellowships towards the creation of science technology innovation, Grant Number JPMJFS2129. This work was supported by Japan Society for the Promotion of Science (JSPS) KAKENHI Grant Numbers JP21H01137. This work was also partially supported by Optical and Near-Infrared Astronomy Inter-University Cooperation Program from the Ministry of Education, Culture, Sports, Science and Technology (MEXT) of Japan. We are grateful to the observation and operating members of Kanata Telescope. The Liverpool Telescope is operated on the island of La Palma by Liverpool John Moores University in the Spanish Observatorio del Roque de los Muchachos of the Instituto de Astrofisica de Canarias with financial support from the UKRI Science and Technology Facilities Council (STFC) (ST/T00147X/1). B. A.-G., S.K., and I.L were funded by the European Union ERC-2022-STG - BOOTES - 101076343. 
The VLBA is an instrument of the National Radio Astronomy Observatory. The National Radio Astronomy Observatory is a facility of the National Science Foundation operated by Associated Universities, Inc.

Views and opinions expressed are however those of the author(s) only and do not necessarily reflect those of the European Union or the European Research Council Executive Agency. Neither the European Union nor the granting authority can be held responsible for them. This research has made use of data from the RoboPol program, a collaboration between Caltech, the University of Crete, IA-FORTH, IUCAA, the MPIfR, and the Nicolaus Copernicus University, which was conducted at Skinakas Observatory in Crete, Greece. D.B. acknowledges support from the European Research Council (ERC) under the Horizon ERC Grants 2021 program under the grant agreement No. 101040021. 

The data in this study include observations made with the Nordic Optical Telescope, owned in collaboration by the University of Turku and Aarhus University, and operated jointly by Aarhus University, the University of Turku and the University of Oslo, representing Denmark, Finland and Norway, the University of Iceland and Stockholm University at the Observatorio del Roque de los Muchachos, La Palma, Spain, of the Instituto de Astrofisica de Canarias. The data presented here were obtained in part with ALFOSC, which is provided by the Instituto de Astrof\'{\i}sica de Andaluc\'{\i}a (IAA) under a joint agreement with the University of Copenhagen and NOT. We acknowledge funding to support our NOT observations from the Finnish Centre for Astronomy with ESO (FINCA), University of Turku, Finland (Academy of Finland grant nr 306531).

\par
\textit{Software:} XSPEC \citep{arnaud96}
\par
\textit{Facilities:} Effelsberg-100m, IXPE, Kanata telescope, KVN, Liverpool Telescope, LX-200, NOT, NuSTAR, Perkins 1.8m telescope, Swift, Skinakas observatory, SMA, SNO, T60

\bibliography{refs}{}
\bibliographystyle{aasjournal}

\appendix

\section{Spectropolarimetric fit}

\label{sec:appA}
\begin{figure*}\gridline{\leftfig{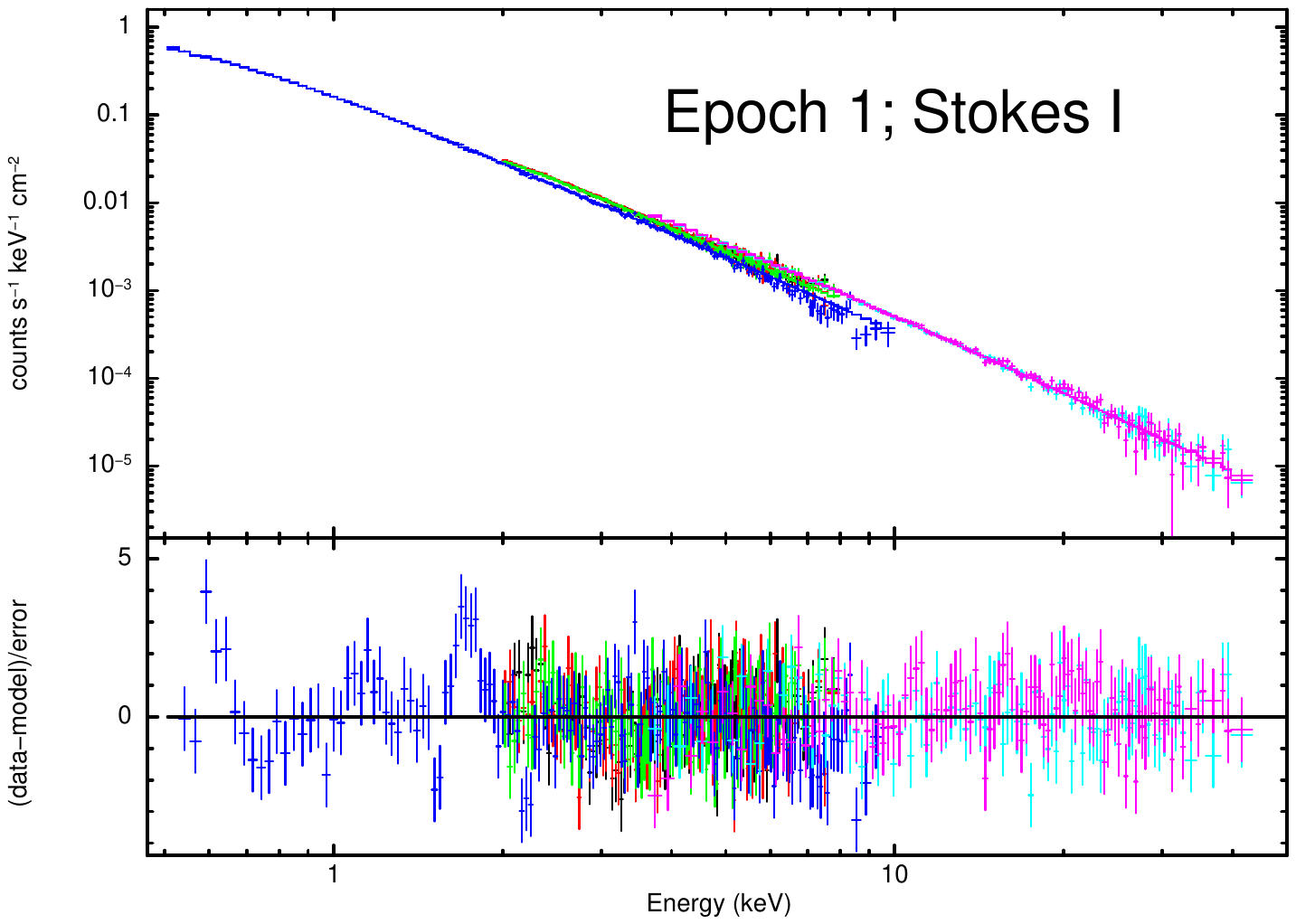}{0.5\textwidth}{(a)}                             \rightfig{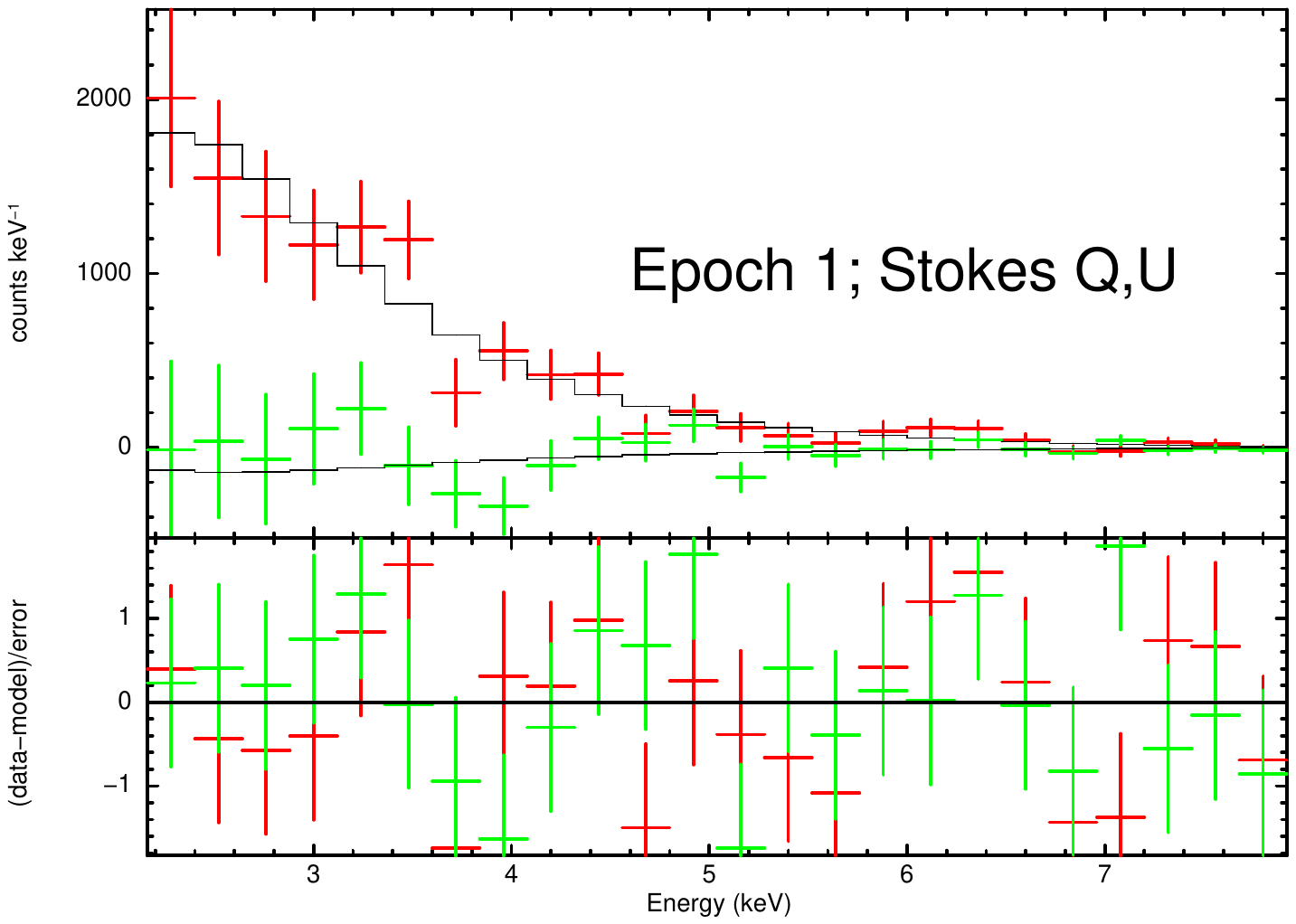}{0.5\textwidth}{(b)}}
\gridline{\leftfig{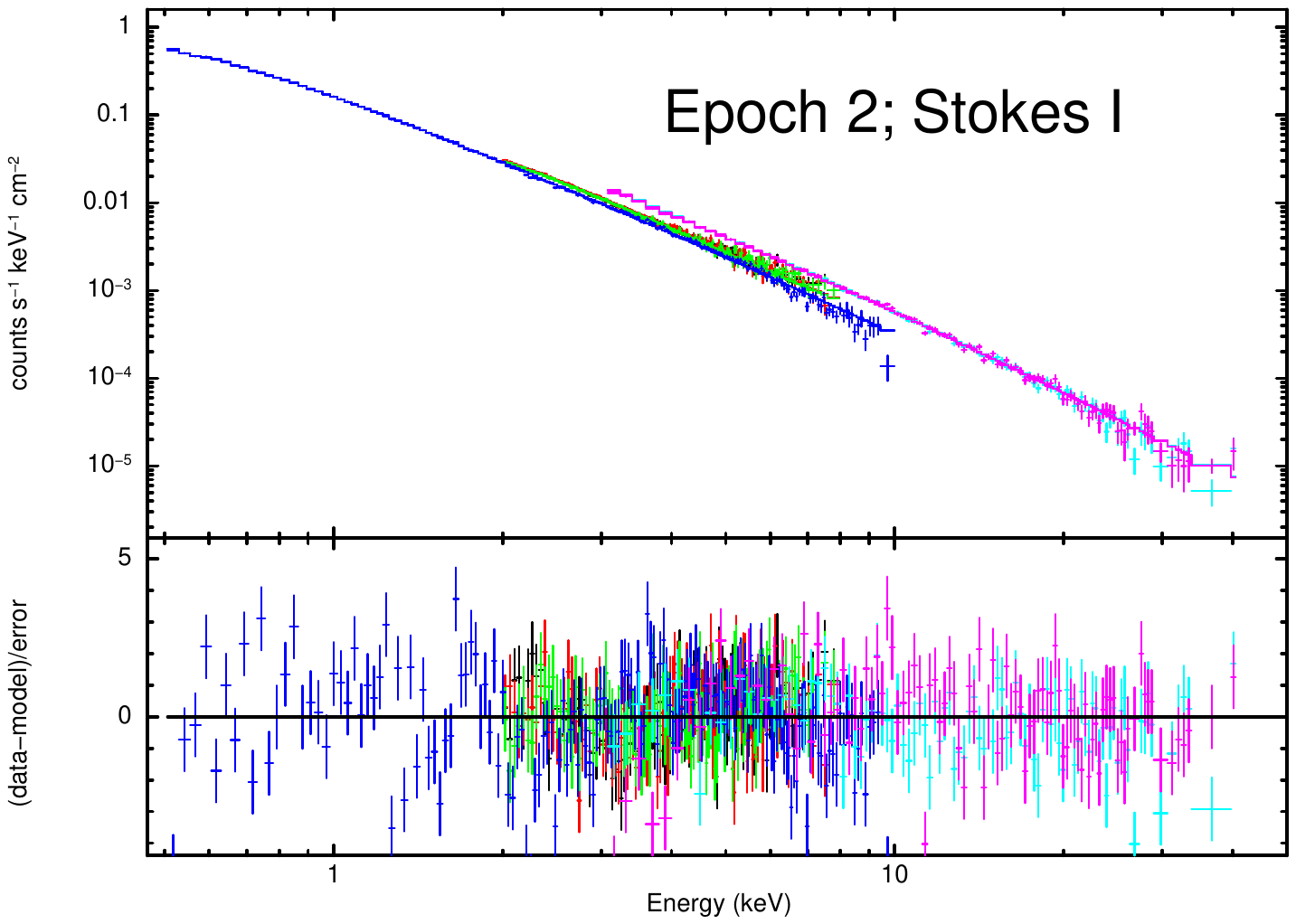}{0.5\textwidth}{(c)}          \rightfig{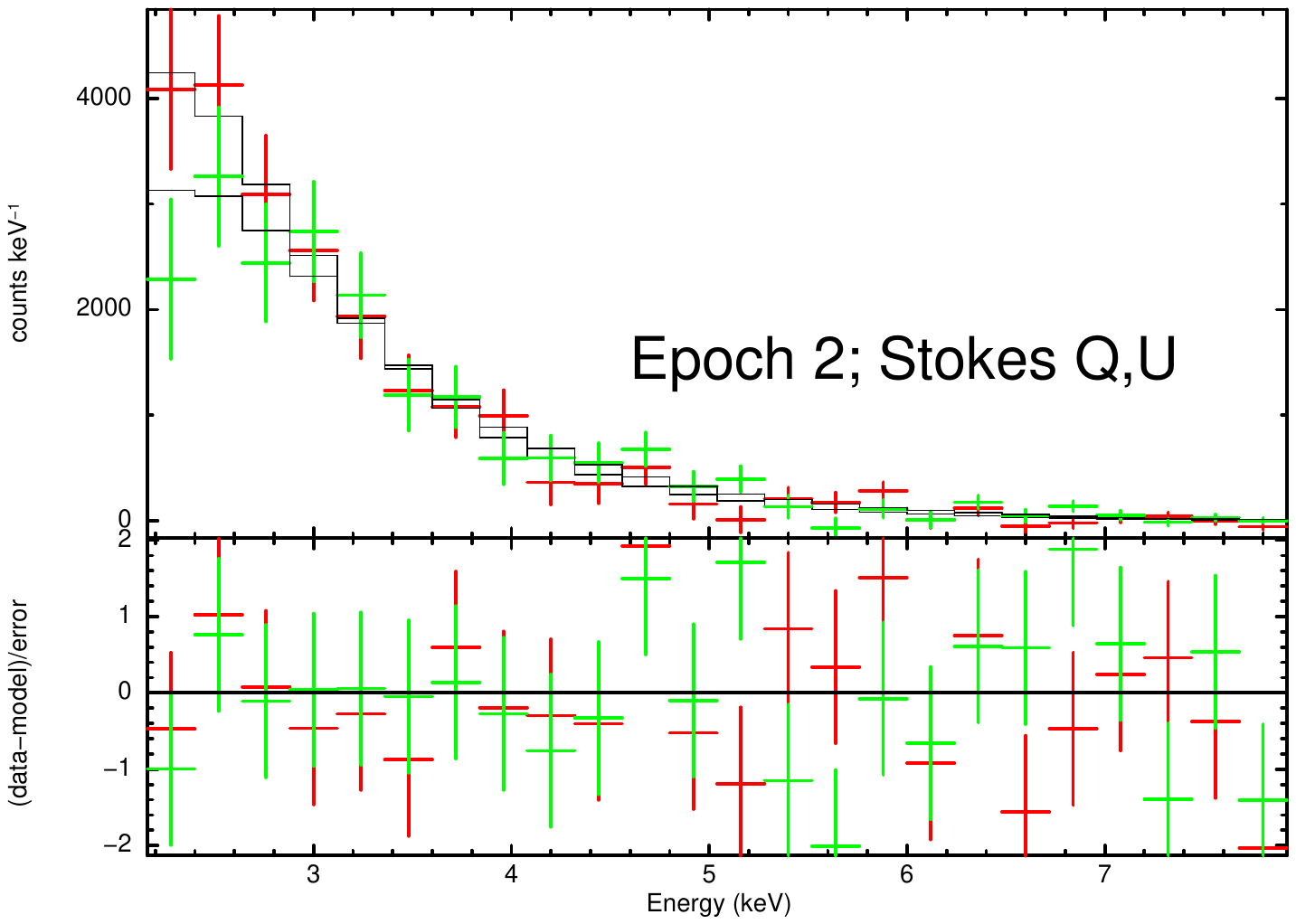}{0.5\textwidth}{(d)}}

\caption{Unfolded I spectra (left) and Q and U spectra (right) for the first (a and b) and second (c and d) {\it IXPE} observations. In the unfolded spectra, dark blue is {\it XMM}, green, red and black are {\it IXPE}, and pink and light blue are {\it NuSTAR}. In the Q and U spectra, red is Q and green is U. Data points are shown as error bars, the model is shown as a solid black step curve.\label{fig:spectra12}}
\end{figure*}

\begin{figure*}
\gridline{\leftfig{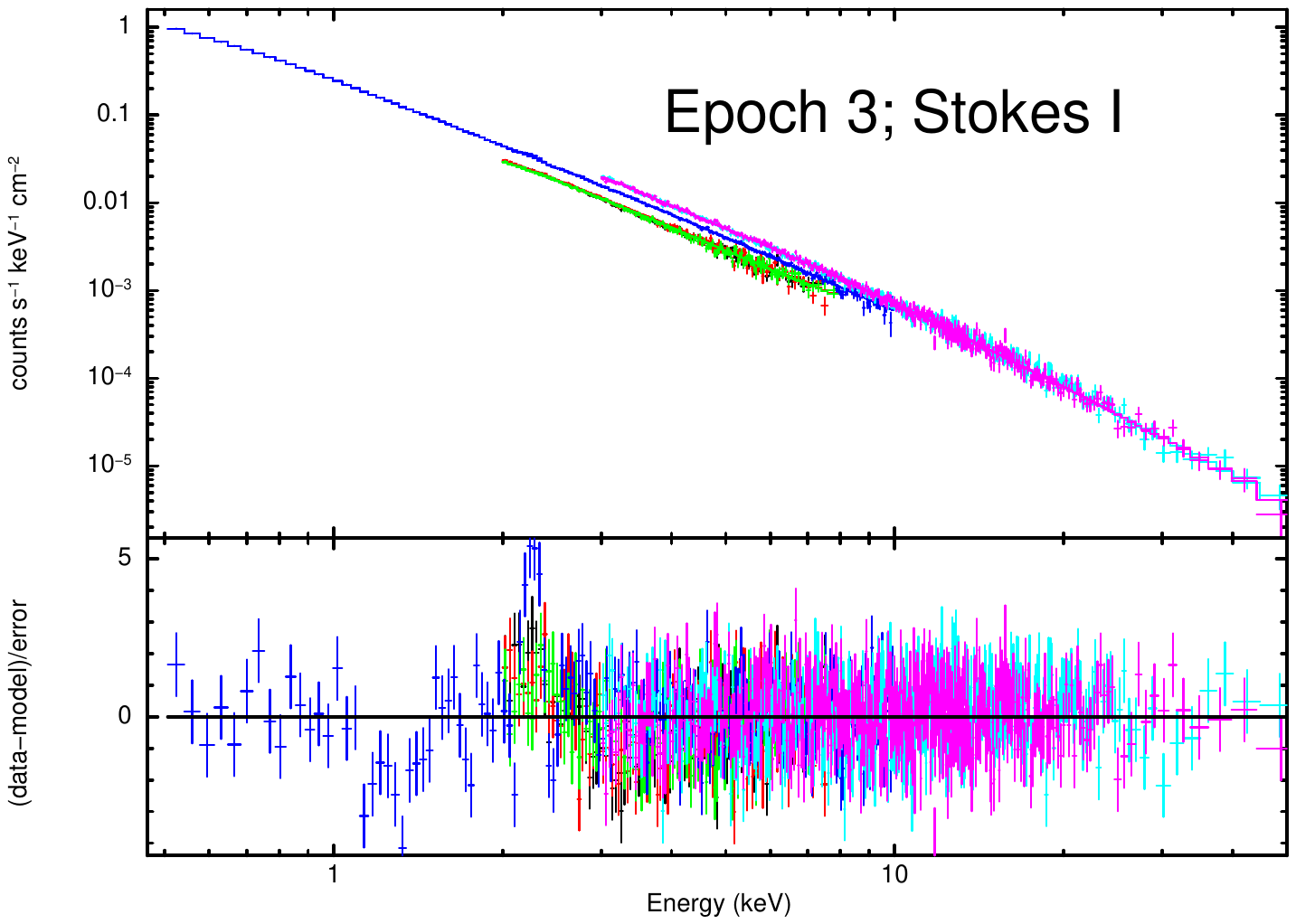}{0.5\textwidth}{(e)}          \rightfig{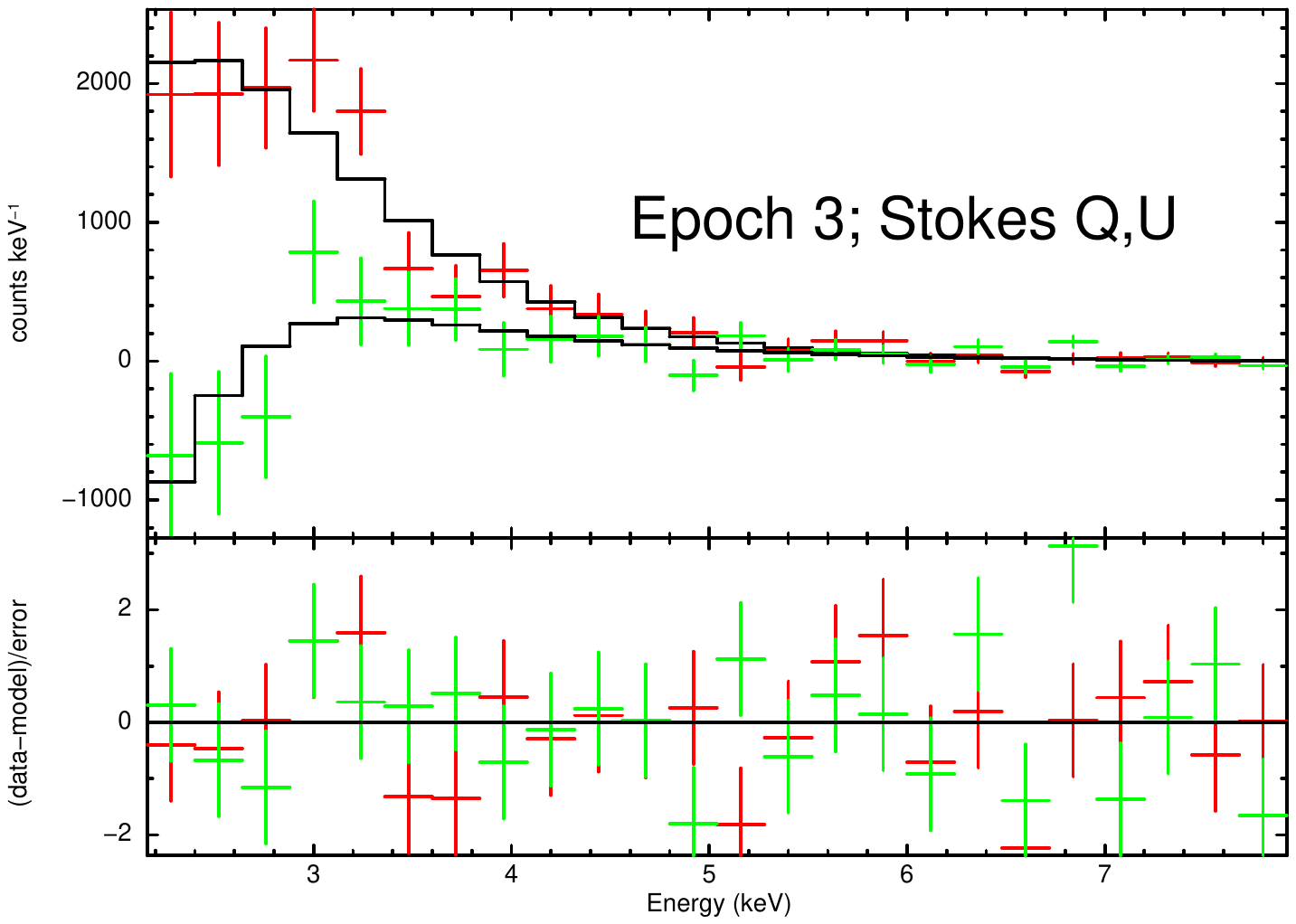}{0.5\textwidth}{(f)}}
\gridline{\leftfig{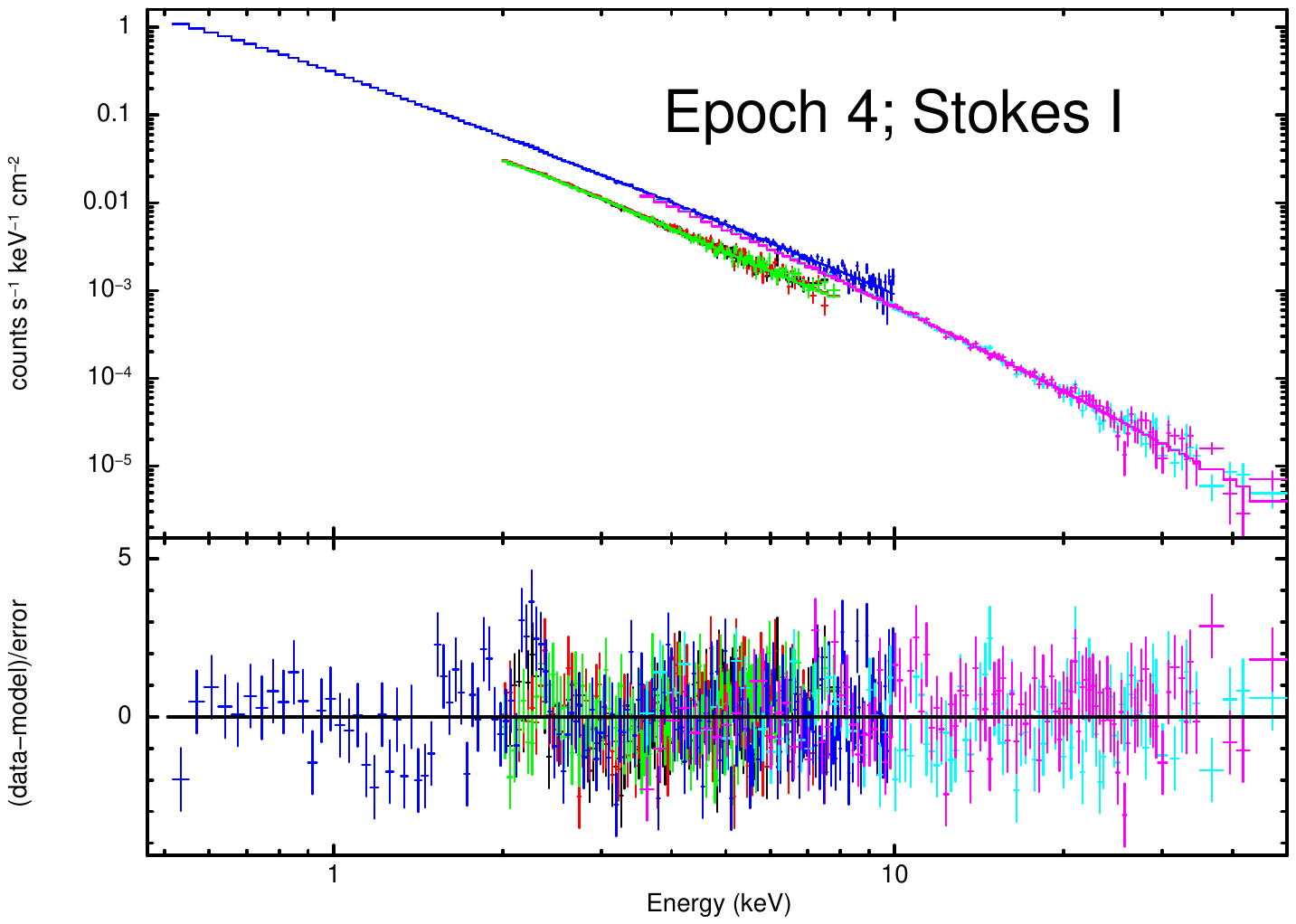}{0.5\textwidth}{(g)}          \rightfig{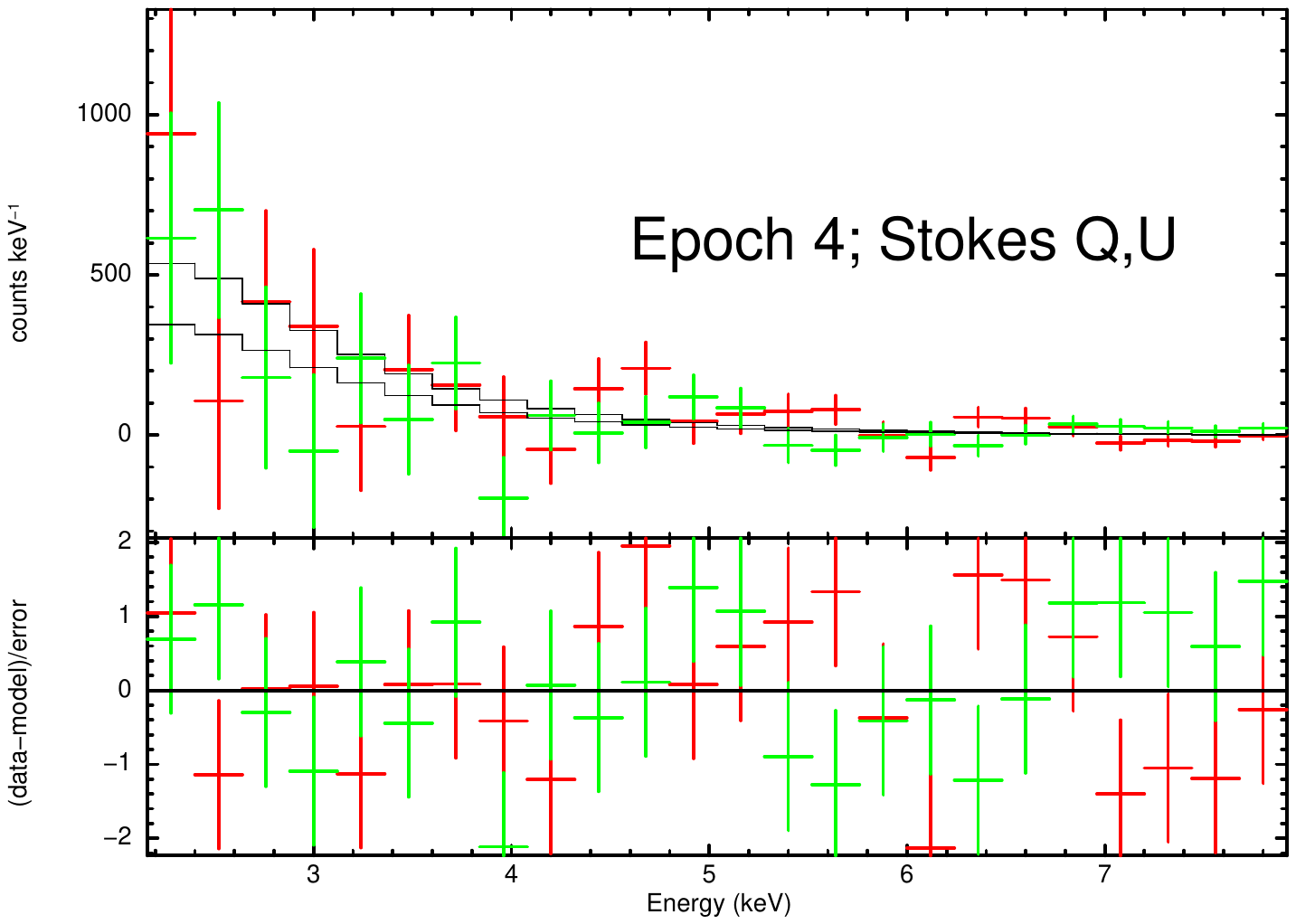}{0.5\textwidth}{(h)}}
\caption{Unfolded I spectra (left) and Q and U spectra (right) for the third (e and f) and fourth (g and h) {\it IXPE} observations. In the unfolded spectra, dark blue is {\it XMM}, green, red and black are {\it IXPE}, and pink and light blue are {\it NuSTAR}. In the Q and U spectra, red is Q and green is U.  Data points are shown as error bars, the model is shown as a solid line.\label{fig:spectra34}}
\end{figure*}

Here we present figures showing the X-ray spectra from {\it IXPE}, {\it XMM-Newton} and {\it NuSTAR}, as well as best-fit log-parabolic models.  The first and second {\it IXPE} observations are shown in Figure \ref{fig:spectra12}, and third and fourth {\it IXPE} observations in Figure \ref{fig:spectra34}. 

%%%%%%%%%%%%%%%%%%%%%%%%%%%%%%%%%%%%%%%%%%%%%%%%%%%%%%%%%%%%%%%%%%%%%%%%%

\section{Multiwavelength polarization observations}
\label{sec:appB}

\begin{figure*}
    \centering
    \includegraphics[width=\textwidth]{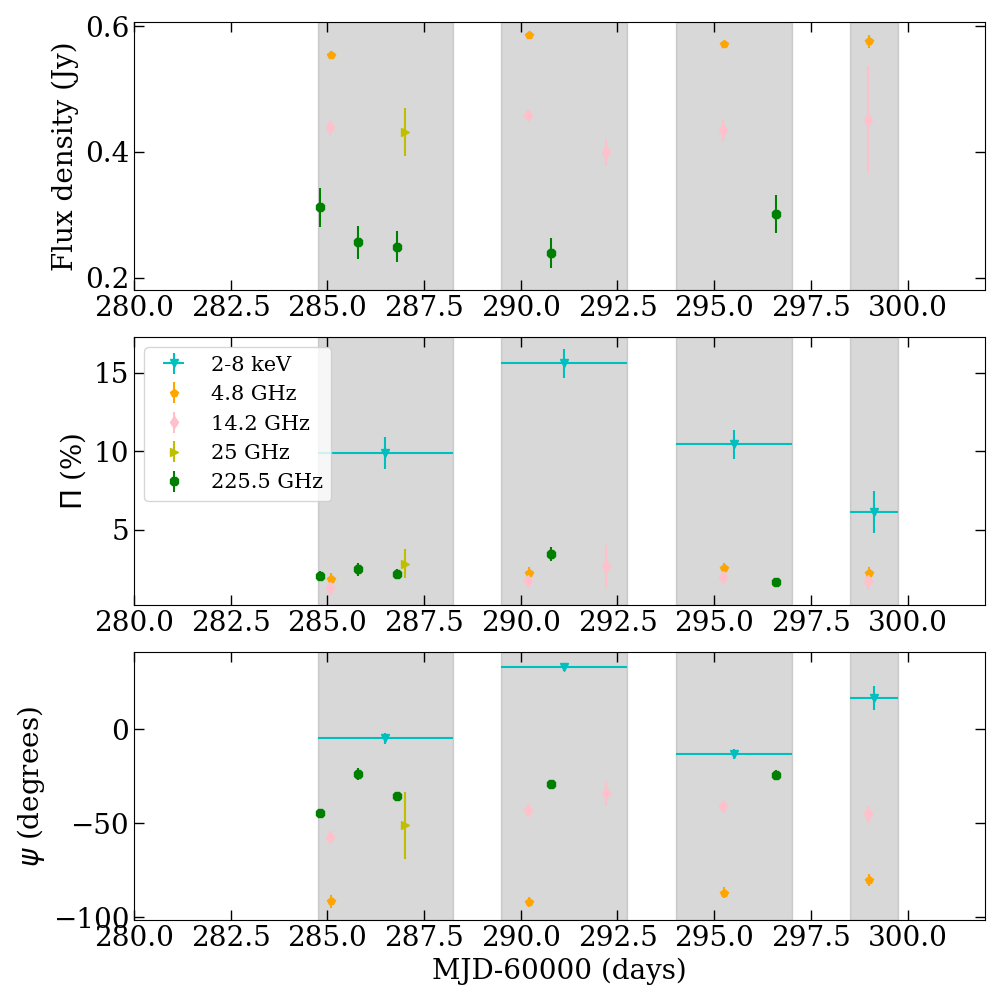}
    \caption{Integrated {\it IXPE} and simultaneous radio polarization observations of Mrk~421. The panels show flux density in Janskys (top), polarization degree in \% (middle), and polarization angle in degrees (bottom). The gray shaded areas mark the duration of the {\it IXPE} observations. The symbols and colors for the different bands are marked in the legend and are the same for all panels.}
    \label{plt:obs_radio}
\end{figure*}

\begin{figure*}
    \centering
    \includegraphics[width=\textwidth]{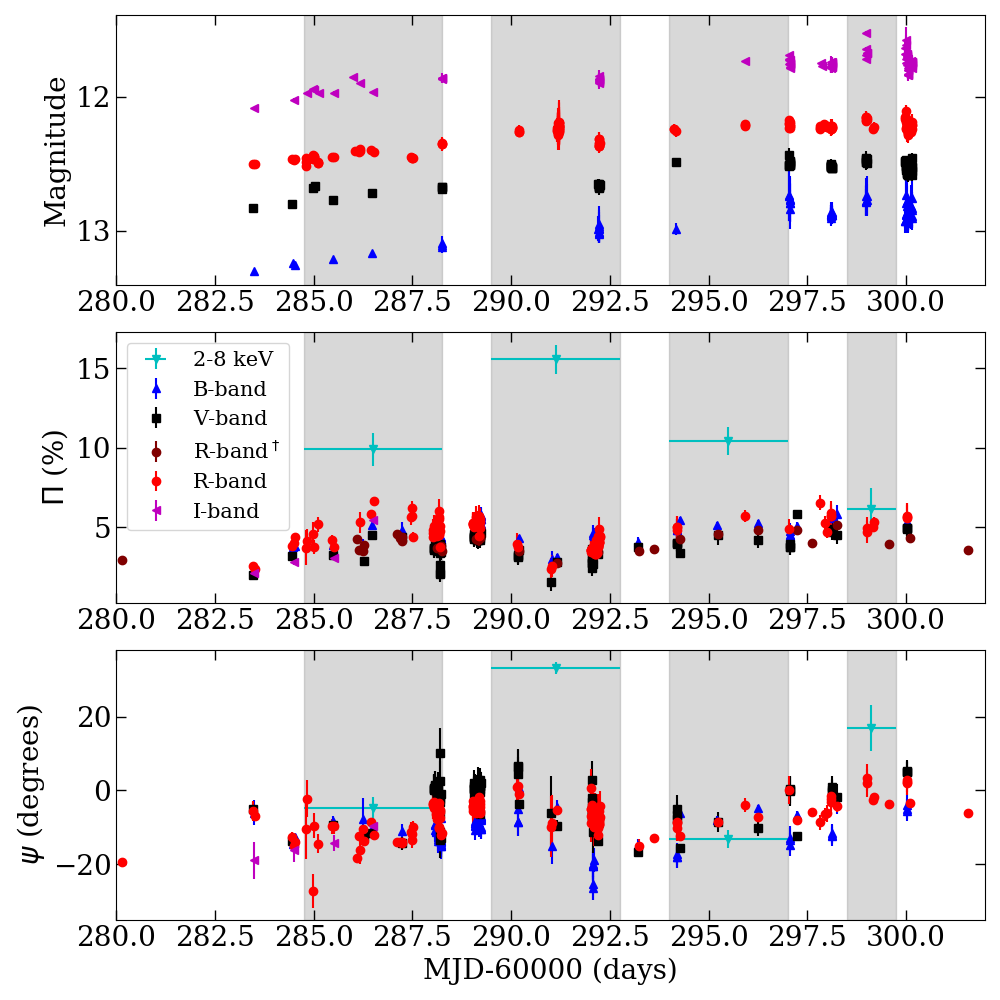}
    \caption{Integrated {\it IXPE} and simultaneous optical polarization observations of Mrk~421. The panels show brightness in magnitudes (top), polarization degree in \% (middle), and polarization angle in degrees (bottom). The gray shaded areas mark the duration of the {\it IXPE} observations. The symbols and colors for the different bands are marked in the legend and are the same for all panels. $^\dagger$ refers to R-band polarization degree that is uncorrected for host-galaxy dilution.}
    \label{plt:obs_optical}
\end{figure*}

Here we present Figure \ref{plt:obs_radio} and Figure \ref{plt:obs_optical}, which show the flux density or magnitude of radio and optical observations (respectively) in comparison with the per-epoch time-averaged polarization degree and angle {\it IXPE} data for this campaign.

\section{Random-walk simulations}\label{sec:RWS}
\label{sec:appC}
\begin{figure*}
    \centering
    \includegraphics[width=\textwidth]{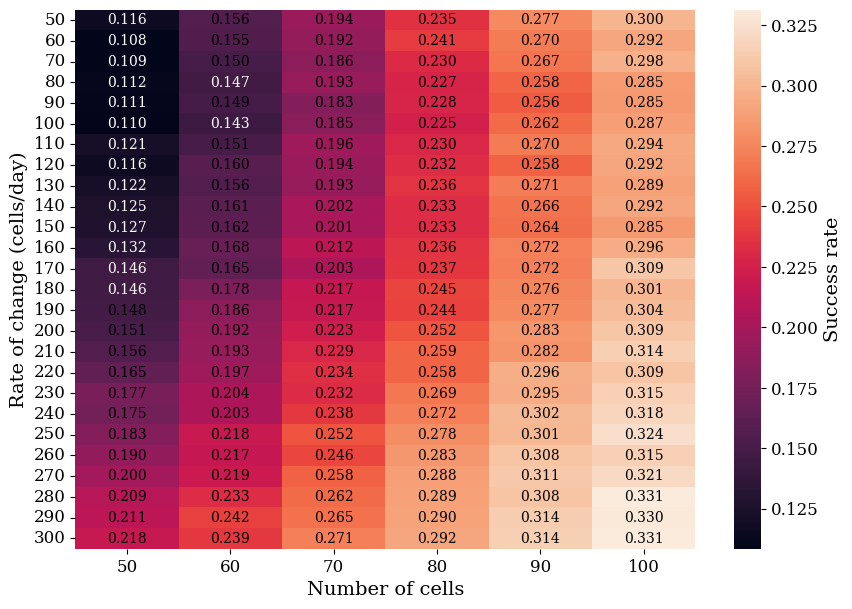}
    \caption{Parameter space of the random-walk simulations. The color bar shows the fraction of successful simulations that reproduce the median polarization degree within 10\% of the observed value.}
    \label{plt:rw_pd}
\end{figure*}

\begin{figure*}
    \centering
    \includegraphics[width=\textwidth]{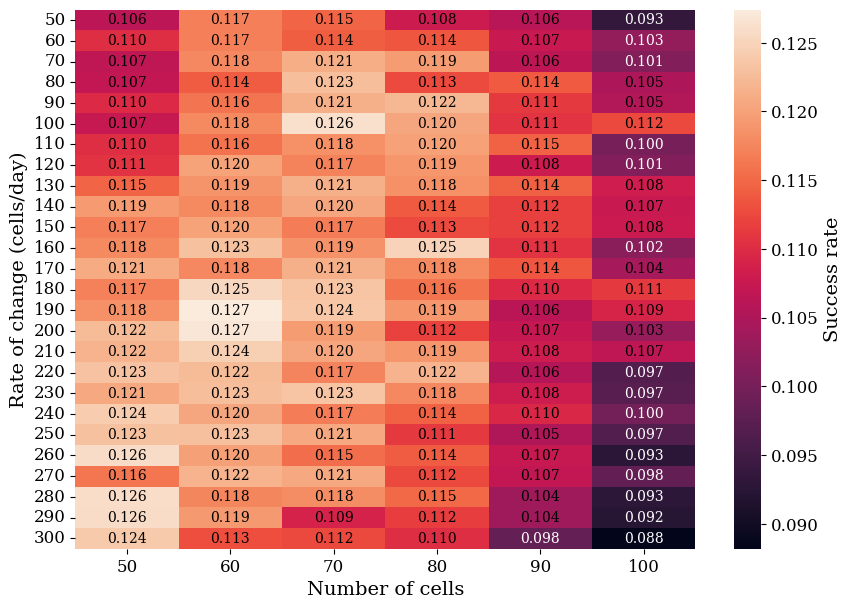}
    \caption{Parameter space of the random-walk simulations. The color bar shows the fraction of successful simulations that reproduce the inter-quartile range of the polarization degree within 10\% of the observed value.}
    \label{plt:rw_pd_iqd}
\end{figure*}

\begin{figure*}
    \centering
    \includegraphics[width=\textwidth]{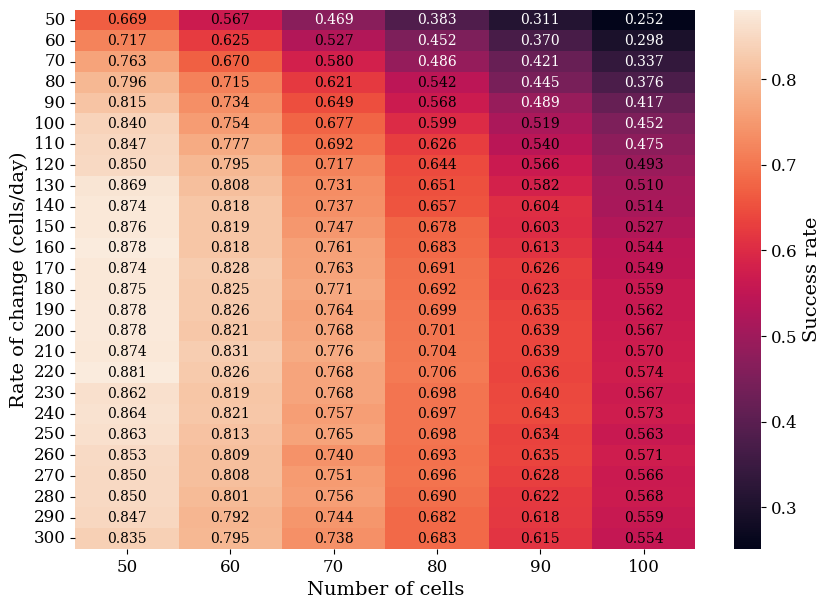}
    \caption{Parameter space of the random-walk simulations. The color bar shows the fraction of successful simulations that reproduce an equal or larger amplitude rotation of the polarization angle compared with the observed value.}
    \label{plt:rw_pa}
\end{figure*}

\begin{figure*}
    \centering
    \includegraphics[width=\textwidth]{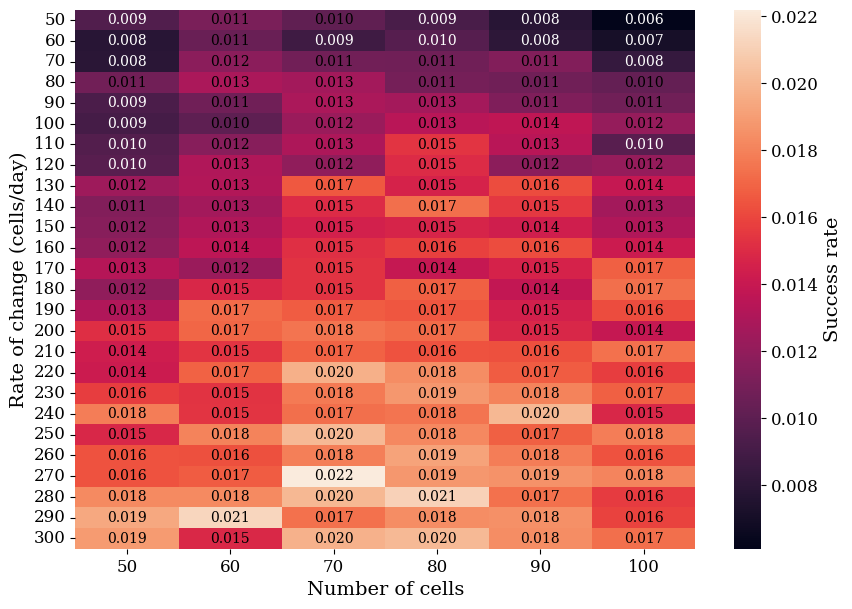}
    \caption{Parameter space of the random-walk simulations. The color bar shows the fraction of successful simulations that reproduce the observed median and inter-quartile range of the polarization degree (within 10\%) and show an equal or larger amplitude rotation of the polarization angle relative to the observed value.}
    \label{plt:rw_all}
\end{figure*}

Here we present Figures \ref{plt:rw_pd}, \ref{plt:rw_pd_iqd}, \ref{plt:rw_pa}, and \ref{plt:rw_all}, which show the success rate of the simulations for the $N_{\rm cell}$, $N_{\rm Var}$ parameter space.

%%%%%%%%%%%%%%%%
\end{document}